\documentclass{aa}
\usepackage{aalongtable}
\usepackage{natbib}
\usepackage{graphics}
\usepackage{txfonts}
\usepackage{ulem}
\usepackage{times}
\usepackage{psfig}
\newfont{\tss}{cmssdc10 scaled 950}

\begin{document}

\title{VLT/X-shooter observations of blue compact galaxies Haro 11 and 
ESO 338-IG 004\thanks{Based on 
observations collected at the European Southern Observatory, Chile, ESO 
programme 60.A-9433(A).}$^,$\thanks{Figures 1, 2 and 3 and Table 1 are available
in electronic form at http://www.aanda.org.}}

\author{N. G.\ Guseva \inst{1,2}    
\and Y. I.\ Izotov \inst{1,2}
\and K. J.\ Fricke \inst{1,3}
\and C.\ Henkel \inst{1,4}}
\offprints{N. G. Guseva, guseva@mao.kiev.ua}
\institute{          Max-Planck-Institut f\"ur Radioastronomie, Auf dem H\"ugel 
                     69, 53121 Bonn, Germany
\and
                     Main Astronomical Observatory,
                     Ukrainian National Academy of Sciences,
                     Zabolotnoho 27, Kyiv 03680,  Ukraine
\and
                     Institut f\"ur Astrophysik, 
                     G\"ottingen Universit\"at, Friedrich-Hund-Platz 1, 
                     37077 G\"ottingen, Germany
\and
                     Astronomy Department, King Abdulaziz University, 
                     P.O. Box 80203, Jeddah, Saudi Arabia
}
\date{Received \hskip 2cm; Accepted}

\abstract
{Strongly star-forming galaxies of subsolar metallicities are typical 
of the high-redshift universe. Here we therefore provide accurate data 
for two low-$z$ analogs, the well-known low-metallicity 
emission-line galaxies Haro 11 and ESO 338-IG 004.
} 
{Our main goal is to derive their spectroscopic 
properties and to examine whether a previously reported near-infrared (NIR) 
excess in Haro 11 can be confirmed.
}  
{On the basis of Very Large Telescope/X-shooter spectroscopic observations
in the wavelength range $\sim$$\lambda\lambda$3000 -- 24000\AA, we use
standard direct methods to 
derive physical conditions and element abundances.
Furthermore, we use X-shooter data together with
{{\sl Spitzer}} observations in the mid-infrared range to attempt to find
hidden star formation.
} 
{We derive interstellar oxygen abundances 
of 12 + log O/H = 8.33 $\pm$ 0.01, 8.10 $\pm$ 0.04, and 7.89 $\pm$ 0.01
in the two H {{\sc ii}} regions B and C of Haro 11 and in ESO 338-IG 004, 
respectively. 
 The observed fluxes of the hydrogen lines 
correspond to the theoretical recombination values after correction for  
extinction with a single value of the extinction coefficient $C$(H$\beta$) 
across the entire wavelength range from the near-ultraviolet to the NIR and 
mid-infrared for each 
of the studied H {{\sc ii}} regions. 
  Thus, we confirm our previous findings obtained for several low-metallicity 
emission-line galaxies (Mrk 59, II Zw 40, Mrk 71, Mrk 996, SBS 0335--052E, 
PHL 293B, and GRB HG 031203)
that the extinction coefficient $C$(H$\beta$) is not higher in the NIR
than in the optical range 
and therefore that there are no emission-line regions contributing to the
line emission in the NIR range, which are hidden in the optical range.
  The agreement between the  
extinction-corrected and CLOUDY-predicted fluxes implies 
that a H {{\sc ii}} region model including only stellar photoionisation 
is able to account for the observed fluxes, in both the optical and NIR ranges.
No additional excitation mechanism such as shocks from stellar 
winds and supernova remnants is needed. 
  All observed spectral energy distributions (SEDs) can be reproduced 
quite well across the whole wavelength range by model SEDs   
except for Haro 11B, where there is a continuum flux excess at 
wavelengths $>$1.6$\mu$m. It is possible that one or more red supergiant 
stars are responsible for the NIR flux excess in Haro 11B.
We find evidence of a luminous blue variable (LBV)
star in Haro 11C.
}
{}
\titlerunning{VLT/X-shooter observations of Haro 11 and ESO 338-IG 004}
\authorrunning{N.G.Guseva et al.}
\maketitle

\section{Introduction \label{intro}}

   Haro 11 ($\equiv$ESO 0350-IG 038) and ESO 338-IG 004 
($\equiv$Tololo 1924--416) are prominent
blue compact galaxies (BCG) with 
$M_B$ = -- 20.0 mag and --18.9 mag,
respectively \citep{BergvallOstlin2002}.
Both exhibit perturbed morphologies, high star-formation rates ($SFRs$),
and multiple-component H$\alpha$ velocities fields
reflecting signatures of merger events 
\citep{OstlinAmram2001}.

   Haro 11 is a massive and luminous BCG with a $SFR$ 
$\sim$18--20 $M_{\odot}$ yr$^{-1}$ \citep{BergvallOstlin2002}
which possesses multiple  
H {\sc ii} regions, called A, B, and C following the 
nomenclature of \citet{Kunth2003}.
\citet{OstlinAmram2001} estimated that the
stellar mass of the galaxy is 10$^{10}$ $M_{\odot}$ 
from the H$\alpha$ velocity field, while   
\citet{BergvallMasegosa2000} derived a total gas mass 
of 2$\times$10$^{9}$ $M_{\odot}$. The high present $SFR$ of 
22$\pm$3 $M_{\odot}$ yr$^{-1}$ in Haro 11 was derived by \citet{Adamo2010}. 
  The galaxy is a Ly$\alpha$ and Ly-continuum emitter \citep{Hayes2007} and
also a luminous IRAS source \citep{BergvallMasegosa2000}
with an IR luminosity of 1.9$\times$10$^{11}$ $L_{\odot}$, which is a 
characteristic property of luminous IR galaxies (LIRGs).
    High star-formation activity in Haro 11 and its global properties
are similar to those in 
high-redshift Lyman break galaxies (LBGs) at $z$ $\sim$ 3 
\citep{Pettini2001} and those in lower-redshift luminous compact 
emission-line galaxies (LCGs) 
from SDSS DR7 \citep{IGT2011} and in green pea galaxies
\citep{Cardamone2009}. 

   ESO 338-IG 004 contains two H {\sc ii} regions in its 
most active part, which is referred to its ``centre'' and ``H {\sc ii} region'' 
\citep{Bergval1985,BergvallOstlin2002}.
The central starburst (``centre'') of the 
galaxy is resolved by {\sl Hubble Space Telescope}
({\sl HST}) images into 
compact star clusters \citep{Meurer1995,OstlinBergval1998}
and young globular clusters \citep{OstlinBergval1998}.
   \citet{Bergval1985} estimated the dynamical mass of the galaxy,
$M_{\rm tot}$$\sim$ 1$\times$10$^{9}$$M_{\odot}$.
\citet{Ostlin2007} derived the dynamical mass 
$M$ = 1.3$\times$10$^{7}$$M_{\odot}$ and an age of $\sim$6 Myr for the brightest 
cluster \#23 in ESO 338-IG 004  from an analysis of the absorption components 
of H$\delta$ and H8 -- H11 Balmer lines.

 \citet{BergvallOstlin2002} obtained a very red colour $V-K$=4.2$\pm$0.8 mag 
for a Haro 11 underlying LSB component. 
On the other hand, \citet{Micheva2010} derived a far less red $V-K$ 
colour of the Haro 11 LSB component of 2.3$\pm$0.2 mag from new deep 
$V$ and $K$ photometry.
   On the basis of high-resolution {\sl HST} imaging of Haro 11 
in eight wavelength bands from the ultraviolet (UV) to $K$, 
\citet{Adamo2010} found about 100 
star clusters with flux excesses at wavelengths $>$ 8000\AA\ with 
respect to the synthetic evolutionary models. This is almost half of the 
whole cluster sample in Haro 11. 
   Moreover, both of the giant H {\sc ii} regions of Haro 11B and C are very 
massive ($\sim$1$\times$10$^7$ $M_{\odot}$) and have a red excess. 

  Based on  ESO 3.6m telescope observations, \citet{Bergval1985} obtained 
an oxygen abundance of 12 + log O/H = 8.08 for the central 
3\arcsec$\times$4\arcsec\ region of ESO 338-IG 004
by applying the direct $T_{\rm e}$-method with the use of the 
emission line [O {\sc iii}]$\lambda$4363\AA. 
An oxygen abundance 12 + log O/H = 7.92, which was derived using  
the $T_{\rm e}$-method,
was obtained by \citet{Masegosa1994} and  \citet{Raimann2000}
from the same spectra acquired by \citet{Terlevich1991}.

Finally, \citet{BergvallOstlin2002} using ESO 1.5-m and 3.6-m spectra 
(obtained in 1983, 1984 and 1986) derived 12 + log O/H = 7.9 and 8.0 for 
Haro 11 and ESO 338-IG 004, respectively, again using the $T_{\rm e}$-method.

  We note, however, that in the case of Haro 11, such a fundamental 
parameter as the
metallicity remains uncertain despite of many previous comprehensive studies. 
This is because the important [O {\sc iii}]$\lambda$4363\AA\ line 
is blended with the [Fe {\sc ii}]$\lambda$4359\AA\ line (this paper) and 
therefore an accurate measurement of the former line's flux was impossible
with previous low-spectral-resolution observations.

  We present here high-quality archival VLT/X-shooter spectroscopic 
observations 
of the two bright knots B and C in Haro 11 and 
 of the brightest part of ESO 338-IG 004
over a wide wavelength range $\sim$$\lambda$3000 -- 24000\AA.
 These new medium-spectral-resolution observations allow us 
to more accurately derive reddening, the physical conditions, 
and the element abundances in the H {\sc ii} regions. 
   Moreover, since cool low-mass stars are a main contributor
to the stellar mass and they emit mainly in the near-infrared (NIR) range,
observations of a very broad wavelength range permit us to  
derive more reliable stellar masses for 
the galaxies and their specific $SFR$s.
  The observations from the UV to NIR wavelength ranges also allow us 
to study whether
any hidden star formation is present in these galaxies.
 We note, however, that in low-metallicity compact galaxies with high $SFR$s
the strongly ionised gaseous emission contributes to the total spectral energy 
distribution (SED) and therefore should be properly taken into account for
the stellar mass determination. 
 
  Thus, the main goals of this paper are 
to obtain more accurate element abundances,
to test spectroscopically 
the flux excess 
in the NIR wavelength range found by \citet{Adamo2010} for Haro 11B and C, 
to search for star formation seen in the NIR range 
but hidden in the optical
range, and to estimate the excitation mechanisms of bright emission lines 
using additionally the {\sl Spitzer} observations of \citet{Wu2008} 
in the mid-infrared (MIR) range.

  In Sect.~\ref{obs}, we describe the observations and data reduction.
  We present the main properties of the galaxies in Sect.~\ref{results}. More 
specifically, the element abundances are presented in subsect.~\ref{abund},
the hidden star formation is discussed in subsect.~\ref{hidden}, 
the H$_2$ emission is considered in subsect.~\ref{s:H2}, the CLOUDY
modelling is discussed in subsect.~\ref{CLOUDY}, the possible presence of 
LBV stars in Haro 11C and Haro 11B in subsect.~\ref{LBV}, and the SED
fitting is described in subsect. \ref{SED}. 
Finally, in Sect.~\ref{concl} we summarize our main results.

\section{Observations \label{obs}}

 New spectra of the blue compact emission-line galaxies 
Haro 11 and ESO 338-IG 004 were obtained 
with the VLT/X-shooter on 2009 August 11 and 12, respectively 
[ESO program 60.A-9433(A)].

\setcounter{figure}{3}
\begin{figure*}[t]
\vspace{0.3cm}
\hspace*{0.0cm}\psfig{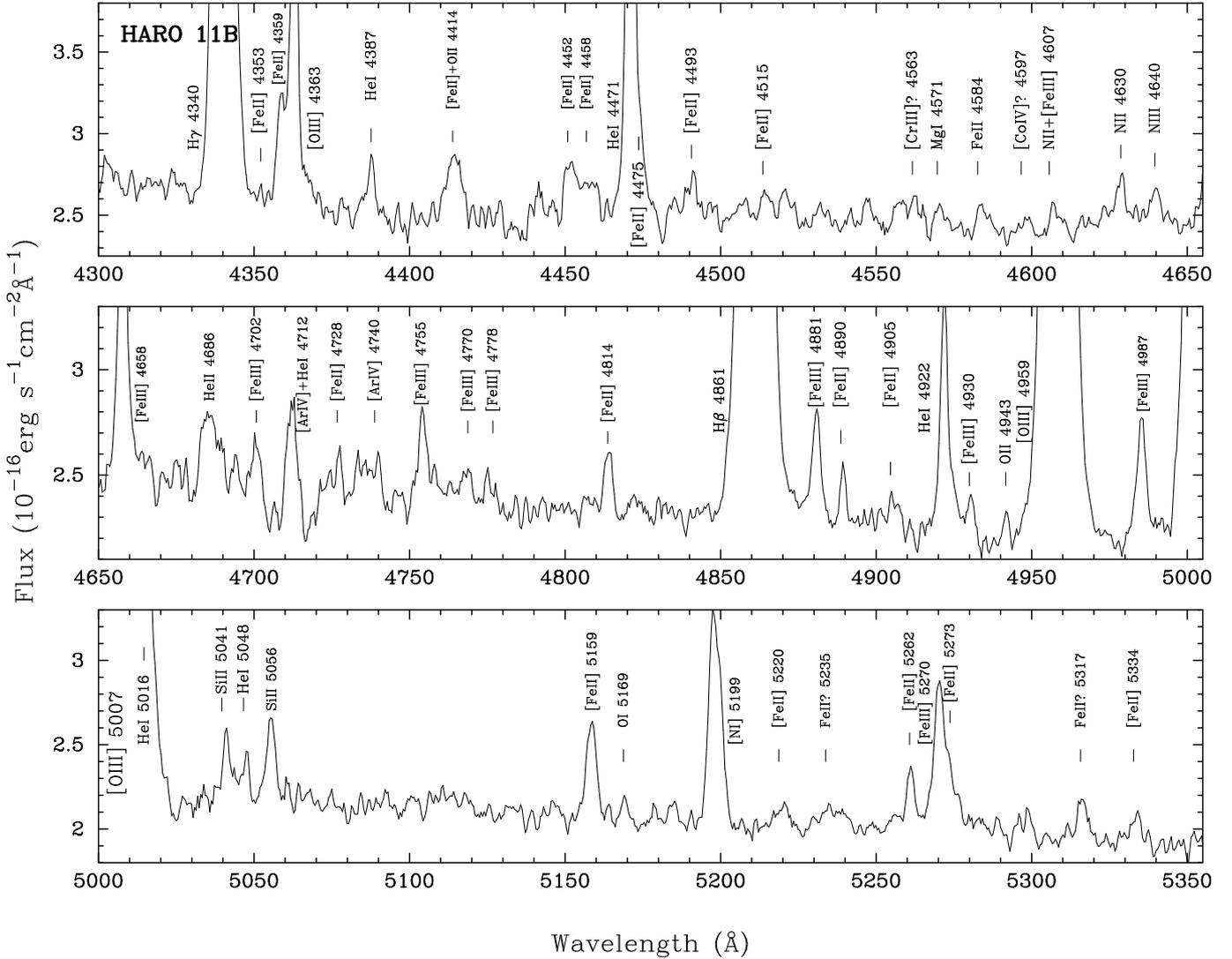}
\caption{Close-up part of the Haro 11B optical spectrum shown in Fig.~\ref{sp_Haro11B}.
}
\label{sp_Haro_11C_detail}
\end{figure*}

\begin{figure*}
\vspace{0.3cm}
\hspace*{1.0cm}\psfig{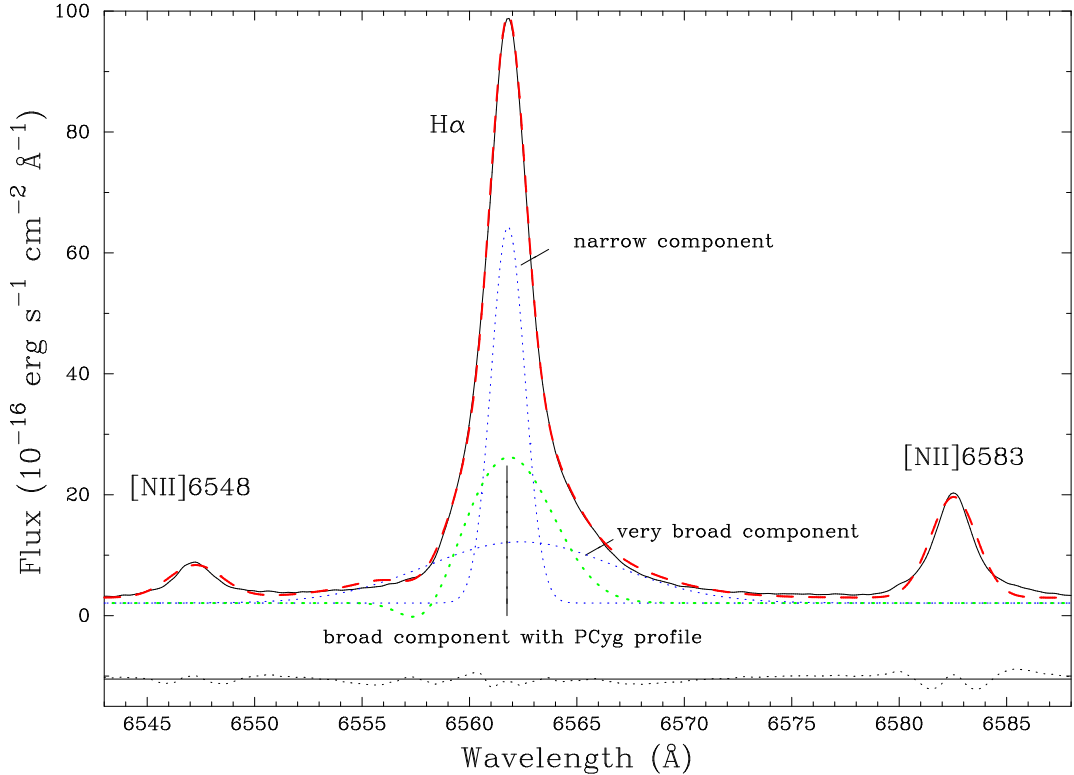}
\caption{Decomposition of the H$\alpha$ emission-line profile in the Haro 11C 
spectrum. Two Gaussian components (blue dotted lines) are likely related to 
nebular emission, a P Cyg profile (green dotted line) is a spectral feature
of an LBV star. The residual is shown by the black dotted line. 
For a better view this is shifted  
below the zero value. 
Observational data are shown by the black solid line and the 
modelled profile, which includes all components, is displayed 
by the red dashed line.
}
\label{d_Ha_2f_NII}
\end{figure*}

   The observations were performed in the 
wavelength range $\sim$$\lambda$3000 -- 24000\AA\ 
using the three-arm echelle X-shooter spectrograph mounted at the UT2 
Cassegrain focus. All observations were obtained at low airmass 
1.016, 1.012, and 1.430 for Haro 11B, Haro 11C, and ESO 338-IG 004, 
respectively, 
thus the effect of the atmospheric dispersion was low.
   The seeings were 0\farcs72 -- 0\farcs75, 0\farcs70 -- 1\farcs32, and
0\farcs61 -- 0\farcs66 for Haro 11B, C, and ESO 338-IG 004 observations, 
respectively.
For each of the UVB, VIS, and NIR arms, the total exposure times were 
800s, 680s, and 600s for Haro 11B, Haro 11C, and ESO 338-IG 004, respectively.
They were broken into two equal subexposures.
  The nodding along the slit was performed according to the scheme AB 
with the object positions A or B differing by 4\arcsec\ along the slit
for each H {\sc ii} region.
  In the UVB arm with wavelength range $\sim$$\lambda$3000 -- 5600\AA, 
a slit of 1\arcsec $\times$ 11\arcsec\ was used.
  In the VIS and NIR arms with wavelength ranges 
$\sim$$\lambda$5500 -- 10200\AA\
and $\sim$$\lambda$10200 -- 24000\AA, respectively,
slits of 0\farcs9 $\times$ 11\arcsec\ were used.
  The binning factors in the UVB and VIS arms 
along the spatial and dispersion axes were 1 and 2, respectively, while
in the NIR arm they were 1 and 1.
 Resolving powers $\lambda$/$\Delta$$\lambda$ of 5100, 8800, and 5100 
characterise the UVB, VIS, and NIR arms, respectively.
  Spectra of thorium-argon (Th-Ar) comparison arcs were 
 used to perform the wavelength calibration of the UVB and VIS arm 
observations. For the wavelength calibration of the NIR spectrum, we used night 
sky emission lines.

   The two-dimensional UVB and VIS spectra were bias subtracted and 
flat-field corrected using IRAF\footnote{IRAF is 
the Image Reduction and Analysis Facility distributed by the 
National Optical Astronomy Observatory, which is operated by the 
Association of Universities for Research in Astronomy (AURA) under 
cooperative agreement with the National Science Foundation (NSF).}.
The NIR dark current frames were subtracted from the NIR flat frames.
The two-dimensional NIR spectra of objects were then
divided by the flat frames to correct for the pixel sensitivity variations.
 Cosmic ray hits of all UVB, VIS, and NIR spectra were removed using the routine 
CRMEDIAN. The remaining hits were later removed manually after background 
subtraction.

  For each of the UVB, VIS, and NIR arms, 
the spectrum at the position B was subtracted from the
spectrum at the position A. This resulted in a frame with
subtracted background. We used the IRAF
software routines IDENTIFY, REIDENTIFY, FITCOORD, and TRANSFORM to 
perform wavelength
calibration and correct for distortion and tilt for each frame. 
The one-dimensional wavelength-calibrated spectra were then extracted from the 
two-dimensional frames using the APALL routine. 
We adopted extraction apertures of  
1\arcsec $\times$ 4\arcsec,  
0\farcs9 $\times$ 4\arcsec, and 
0\farcs9 $\times$ 4\arcsec\ for the UVB, VIS, and NIR 
spectra, respectively.
Before extraction, the spectra at the positions A and B in the two-dimensional
background-subtracted frames were carefully aligned with the routine ROTATE 
and co-added. The spectrophotometric standard star Feige 110 was used 
to perform the flux calibration. Its spectra in all arms were obtained 
with a wide slit of 5\arcsec$\times$11\arcsec.

  The resulting flux-calibrated and redshift-corrected UVB, VIS, and
NIR spectra of Haro 11B, C, and ESO 338-IG 004 are shown in 
Figs.~\ref{sp_Haro11B},~\ref{sp_Haro_11C}, and ~\ref{sp_E338} 
(available only in the online edition).

\section{Results \label{results}}

\subsection{Element abundances \label{abund}}

We derived element abundances from emission-line fluxes 
using a classical semi-empirical method. 
 The fluxes in all spectra were 
measured using the IRAF SPLOT routine.
 The line flux errors 
include statistical errors derived with SPLOT
from non-flux-calibrated spectra, in addition to errors introduced
by the absolute flux calibration, which we
set to 1\% of the line fluxes, according to the uncertainties in the 
absolute fluxes of relatively bright standard stars 
\citep{Oke1990,Colina1994,Bohlin1996,IT04a}. 
These errors were propagated into the calculation of 
the electron temperatures, the electron number densities, and the ionic and 
total element abundances following the prescription of \citet{Guseva2011}. 

The extinction coefficient 
$C$(H$\beta$) and equivalent widths of the hydrogen absorption lines
EW(abs) are calculated by simultaneously minimising the deviations 
in the corrected fluxes $I(\lambda)$/$I$(H$\beta$) of all hydrogen Balmer lines
from their theoretical recombination values.
The fluxes were corrected for both extinction, using the reddening curve
of \citet{C89}, and underlying
hydrogen stellar absorption \citep{ITL94}.
  The derived $C$(H$\beta$)
was applied to correct all emission-line fluxes in the
entire wavelength range $\lambda$$\lambda$3000 -- 24000\AA.
  The extinction-corrected relative 
fluxes $I$($\lambda$)/$I$(H$\beta$)$\times$100 of the lines, 
the extinction coefficient
$C$(H$\beta$), the equivalent width of the H$\beta$ emission line 
EW(H$\beta$), the H$\beta$ observed flux $F$(H$\beta$), and the 
equivalent width of the 
underlying hydrogen absorption lines EW(abs) are given in Table \ref{Int}
(available only in the online edition).
  We note that the fluxes of the Balmer hydrogen emission lines corrected for 
extinction and underlying hydrogen absorption (Cols., 3, 5, and 7 in 
Table \ref{Int}) 
are close to the theoretical recombination values of \citet{HS87} 
(column 8 of the Table 1), suggesting 
that the extinction coefficient $C$(H$\beta$) is reliably derived.

  The physical conditions, and the ionic and total heavy element 
abundances in the 
ISM of the galaxies were derived following \citet{ISMGT2006} 
(Table \ref{tab2}). 
   In particular, applying the direct $T_{\rm e}$-method, we derived
the electron temperature $T_{\rm e}$(O~{\sc iii}) from the 
[O~{\sc iii}] $\lambda$4363/($\lambda$4959 + $\lambda$5007)
emission-line ratio. 
   The electron temperatures $T_{\rm e}$(O~{\sc ii}) and
$T_{\rm e}$(S {\sc iii}) were derived from the empirical relations 
based on the photoionisation models of H {\sc ii} 
regions \citep{ISMGT2006}.
    The electron number densities  
$N_{\rm e}$(S~{\sc ii}) were obtained from the 
[S~{\sc ii}]$\lambda$6717/$\lambda$6731 emission-line ratios.
   The electron temperatures $T_{\rm e}$(O~{\sc iii}), 
$T_{\rm e}$(O~{\sc ii}), and $T_{\rm e}$(S {\sc iii}),   
the electron number densities $N_{\rm e}$(S~{\sc ii}), 
the ionisation correction factors $ICF$s, and
the ionic and total O, N, Ne, S, Ar, Fe, and Cl abundances derived from the
forbidden emission lines are given in Table \ref{tab2}.   

 We derive 12+logO/H = 8.33$\pm$0.01 and 8.10$\pm$0.04 
for Haro 11B and C, respectively, 
which is appreciably higher than the oxygen abundance 12+logO/H = 7.9 
obtained by \citet{BergvallOstlin2002} in an aperture 
4\arcsec$\times$4\arcsec\ for the brightest region of Haro 11.
  We also obtain a higher extinction $C$(H$\beta$)=0.74 for Haro 11B than
$C$(H$\beta$)=0.27 for Haro 11C. It can indeed be seen in the {\sl HST} images 
obtained by \citet{Adamo2010} that Haro 11B 
is more obscured by  dust clouds than Haro 11C. \citet{Adamo2010} also 
derived different extinctions of 
$E$($B-V$)= 0.38 and 0.06 (or $C$(H$\beta$) = 0.56 and 0.09) for Haro 11B 
and Haro 11C, respectively. 
  For comparison, \citet{Hayes2007} derived $E$($B-V$)=0.42 and 0.48 
(or $C$(H$\beta$)=0.62 and 0.70) and 
\citet{Vader1993} derived $E$($B-V$)=0.41 and 0.39 
for Haro 11B and C, respectively.  

   The differences in oxygen abundance of Haro 11
are likely not solely due to differences in the extinction but 
may also be caused by blending of the  
[O {\sc iii}]$\lambda$4363\AA\ emission line 
with the [Fe {\sc ii}]$\lambda$4359\AA\ one
in previous low-spectral-resolution observations. 
  The higher-spectral-resolution 
X-shooter observations have allowed to resolve this blend.
In Table~\ref{Int} we can clearly see, that
the fluxes of the [Fe {\sc ii}]$\lambda$4359\AA\ emission line 
in Haro 11B and Haro 11C are, respectively, 27\% and 151\% of 
the [O {\sc iii}]$\lambda$4363\AA\ emission-line fluxes.
 
  The oxygen abundance 12+logO/H = 7.89$\pm$0.01 of ESO 338-IG 004 is  
slightly lower than the
previous determinations of 12 + log O/H = 7.92 
\citep{Masegosa1994,Raimann2000}, 8.08 \citep{Bergval1985}, 
8.0 \citep{BergvallOstlin2002} and 7.99 \citep{GISFHP2011}, where the 
direct $T_{\rm e}$-method was also used.
  We note that our determination of
$C$(H$\beta$) = 0.31 (see Table~\ref{Int}) is 
larger than the $C$(H$\beta$) in the range 0.0 -- 0.05 derived
by \citet{Bergval1985}, \citet{Raimann2000}, and \citet{GISFHP2011}, 
and $E$($B-V$)$<$0.1 of
\citet{BergvallOstlin2002}. 
  From the fits of model SEDs to the SEDs derived from 
{\sl HST} images with different filters, \citet{OstlinHayes2009} derived
a low value of $C$(H$\beta$)$\sim$0.1 for ESO 338-IG 004.
  These differences may be caused by the oxygen 
abundances having been derived
by different authors for different regions in ESO 338-IG 004. To check this
possibility, we extracted separate spectra in addition to the integrated 
spectrum, for the ``centre'' and ``H {\sc ii} region''.
 For the centre, we derived 12+log O/H = 7.82$\pm$0.01, which is lower than that
for the integrated spectrum, but for the H {\sc ii} region we obtain a higher
value of 8.00$\pm$0.01, which is more consistent with some of the previous 
observations.

  The abundance ratios of the different species 
log N/O, log Ne/O, log S/O, log Ar/O, log Fe/O, and log Cl/O 
(Table~\ref{tab2}) obtained for Haro 11B, C, and ESO 338-IG 004
are in the range of the ratios obtained for
different samples of blue compact galaxies,
such as the HeBCD sample of 
\citet{ISGT2004} and \citet{IT04a} used to determine the primordial
He abundance (consisting of more than 100 emission-line galaxies), the 
sample of $\sim$300 emission-line galaxies from the SDSS DR3 of 
\citet{ISMGT2006},
and the VLT sample of 121 low-metallicity emission-line galaxies
studied by \citet{GISFHP2011}. 
  The high values of log N/O
in Haro 11B (--0.92, Table \ref{tab2}) and in Haro 11C (--0.79)  
agree with the value of log N/O =--0.7 derived by
\citet{BergvallOstlin2002} for Haro 11. 
   Our values of log N/O, log Ne/O, and log S/O for ESO 338-IG 004 
(Table \ref{tab2}) are consistent with the data of \citet{Masegosa1994} 
(log N/O = --1.40$\pm$0.01 and log Ne/O = --0.75$\pm$0.10),
 \citet{Raimann2000} (log N/O = --1.57),
\citet{Bergval1985} (log N/O = --1.18, and log S/O = --1.63), 
\citet{BergvallOstlin2002} (log N/O = --1.5), and \citet{GISFHP2011}
(log N/O = --1.34, log Ne/O = --0.78, and log S/O = --1.60).

\setcounter{table}{1}


\begin{table}[t]
\caption{Physical conditions and element abundances \label{tab2}}
\begin{tabular}{lccc} \hline\hline
Galaxy       &Haro 11B & Haro 11C& ESO 338-IG 004 \\ \hline
Property     &&Value&         \\ \hline 

$T_{\rm e}$(O {\sc iii}), K                  &10144$\pm$70 & 11362$\pm$449 & 14771$\pm$137  \\
$T_{\rm e}$(O {\sc ii}), K                   &10099$\pm$77 & 11552$\pm$500 & 13786$\pm$160 \\
$T_{\rm e}$(S {\sc iii}), K                  &10110$\pm$58 & 11209$\pm$373 & 14397$\pm$114 \\
$N_{\rm e}$(S {\sc ii}), cm${-3}$                      &10$\pm$10 & 10$\pm$10&142$\pm$36  \\ \\
O$^+$/H$^+$, ($\times$10$^5$)     &8.38$\pm$0.28  &5.21$\pm$0.76  &1.66$\pm$0.06 \\
O$^{2+}$/H$^+$, ($\times$10$^5$)   &12.78$\pm$0.33 &7.48$\pm$0.91  &5.87$\pm$0.16  \\
O$^{3+}$/H$^+$, ($\times$10$^5$)   &0.22$\pm$0.02  &~~~...~~~      &0.19$\pm$0.01 \\
O/H, ($\times$10$^5$)             &21.38$\pm$0.43 &12.69$\pm$1.19 &7.73$\pm$0.17 \\
12+log O/H                        &8.33$\pm$0.01  &8.10$\pm$0.04  &7.89$\pm$0.01 \\ \\
N$^{+}$/H$^+$, ($\times$10$^6$)    &8.87$\pm$0.29  &7.48$\pm$1.13 &0.62$\pm$0.03 \\
$ICF$(N)$^{\rm a}$                      &     2.93      &     2.72     &     4.54    \\
N/H, ($\times$10$^6$)             &25.99$\pm$1.06 &20.37$\pm$3.94 &2.82$\pm$0.16 \\
log N/O                         &--0.92$\pm$0.02~~&--0.79$\pm$0.09~~ &--1.44$\pm$0.03~~\\ \\
Ne$^{2+}$/H$^+$, ($\times$10$^5$)  &3.10$\pm$0.09 &2.09$\pm$0.28 &1.34$\pm$0.04 \\
$ICF$(Ne)$^{\rm a}$                     &      1.49    &      1.42    &      1.13     \\
Ne/H, ($\times$10$^5$)            &4.62$\pm$0.17 &2.97$\pm$0.49 &1.52$\pm$0.05\\
log Ne/O                       &--0.66$\pm$0.02~~&--0.63$\pm$0.08~~ &--0.71$\pm$0.02~~\\ \\
S$^{+}$/H$^+$, ($\times$10$^6$)    &0.57$\pm$0.01 &0.68$\pm$0.06 &0.22$\pm$0.01  \\
S$^{2+}$/H$^+$, ($\times$10$^6$)   &2.39$\pm$0.08 &1.48$\pm$0.31 &0.77$\pm$0.04  \\
$ICF$(S)$^{\rm a}$                      &      1.07    &      1.06    &      1.23    \\
S/H, ($\times$10$^6$)             &3.17$\pm$0.10 &2.29$\pm$0.35 &1.22$\pm$0.05  \\
log S/O                        &--1.83$\pm$0.02~~&--1.74$\pm$0.08~~ &--1.80$\pm$0.02~~\\ \\
Ar$^{2+}$/H$^+$, ($\times$10$^7$)  &6.23$\pm$0.14 &5.59$\pm$0.44 &2.42$\pm$0.06  \\
Ar$^{3+}$/H$^+$, ($\times$10$^7$)  &0.58$\pm$0.05 &~~~...~~~ &0.89$\pm$0.07  \\
$ICF$(Ar)$^{\rm a}$                     &      1.07    &      1.08    &      1.11     \\
Ar/H, ($\times$10$^7$)            &6.69$\pm$0.16 &6.02$\pm$0.52 &2.69$\pm$0.11  \\
log Ar/O                       &--2.50$\pm$0.01~~&--2.32$\pm$0.06~~ &--2.46$\pm$0.02~~\\ \\
$[$Fe {\sc iii}$]$ $\lambda$4658: \\
Fe$^{2+}$/H$^+$, ($\times$10$^6$)&0.99$\pm$0.03&1.32$\pm$0.17 &0.24$\pm$0.01  \\
$ICF$(Fe)$^{\rm a}$                         &      3.73   &      3.48    &      6.17     \\
Fe/H, ($\times$10$^6$)          &3.69$\pm$0.16&4.61$\pm$0.88 &1.49$\pm$0.10  \\
log Fe/O                   &--1.76$\pm$0.02~~&--1.44$\pm$0.09~~ &--1.71$\pm$0.03~~\\ \\
$[$Fe {\sc iii}$]$ $\lambda$4986: \\
Fe$^{2+}$/H$^+$, ($\times$10$^6$)&0.35$\pm$0.02&0.51$\pm$0.09 &0.29$\pm$0.01  \\
$ICF$(Fe)$^{\rm a}$                         &      3.73   &      3.48    &      6.17     \\
Fe/H, ($\times$10$^6$)          &1.30$\pm$0.07&1.78$\pm$0.42 & 1.77$\pm$0.11 \\
log Fe/O                   &--2.21$\pm$0.02~~&--1.85$\pm$0.11~~ &--1.64$\pm$0.03~~\\ \\
Cl$^{2+}$/H$^+$, ($\times$10$^7$)      &0.35$\pm$0.02&0.71$\pm$0.12 &~~~...~~~  \\
$ICF$(Cl)$^{\rm a}$                         &      1.13   &      1.17    &~~~...~~~        \\
Cl/H, ($\times$10$^7$)                &0.39$\pm$0.02&0.84$\pm$0.23 &~~~...~~~  \\
log Cl/O                          &--3.73$\pm$0.03~~&--3.18$\pm$0.13~~ &~~~...~~~\\
\hline
\end{tabular}

\smallskip

$^{\rm a}$ Ionisation correction factor.
\end{table}


\begin{table} [t]
\caption{Stellar masses and ages of young stellar populations
 \label{comp}}
\begin{tabular}{lccc} \hline 
ID& Mass($M_{\odot}$) & Age(Myr)& $C$(H$\beta$) \\ \hline 
{} & \multicolumn{3}{c}{this paper$^{\rm a}$}   \\ 
  \cline{2-4} 
Haro 11B  & 1.26$\times$10$^8$ & 2.9 & 0.74 \\
Haro 11C  & 2.05$\times$10$^8$ & 4.2 & 0.27 \\
ESO 338-IG 004 & 0.56$\times$10$^8$ & 1.9 & 0.37 \\ \cline{2-4}
 {} & \multicolumn{3}{c}{\citet{Adamo2010}$^{\rm b}$}  \\ \cline{2-4}
Haro 11B  & 8.35$\times$10$^6$ & 3.5 & 0.56 \\
Haro 11C  & 1.36$\times$10$^7$ & 9.5 & 0.09 \\ \cline{2-4}
 {} & \multicolumn{3}{c}{\citet{Ostlin2007}$^{\rm c}$}  \\ \cline{2-4}
ESO 338-IG 004 & 1.3$\times$10$^7$ & 6.3 & ... \\ \hline
\end{tabular}

$^{\rm a}$VLT/X-shooter data.

$^{\rm b}${\sl HST} photometric data.

$^{\rm c}$VLT/UVES data for the brightest cluster \#23.
\end{table}



\subsection{Hidden star formation\label{hidden}}

   The spectra of the studied galaxies were obtained 
simultaneously over the entire optical and near-infrared wavelength  
ranges. This enables us to directly compare the
extinctions derived for the optical and NIR spectra.

\begin{figure}
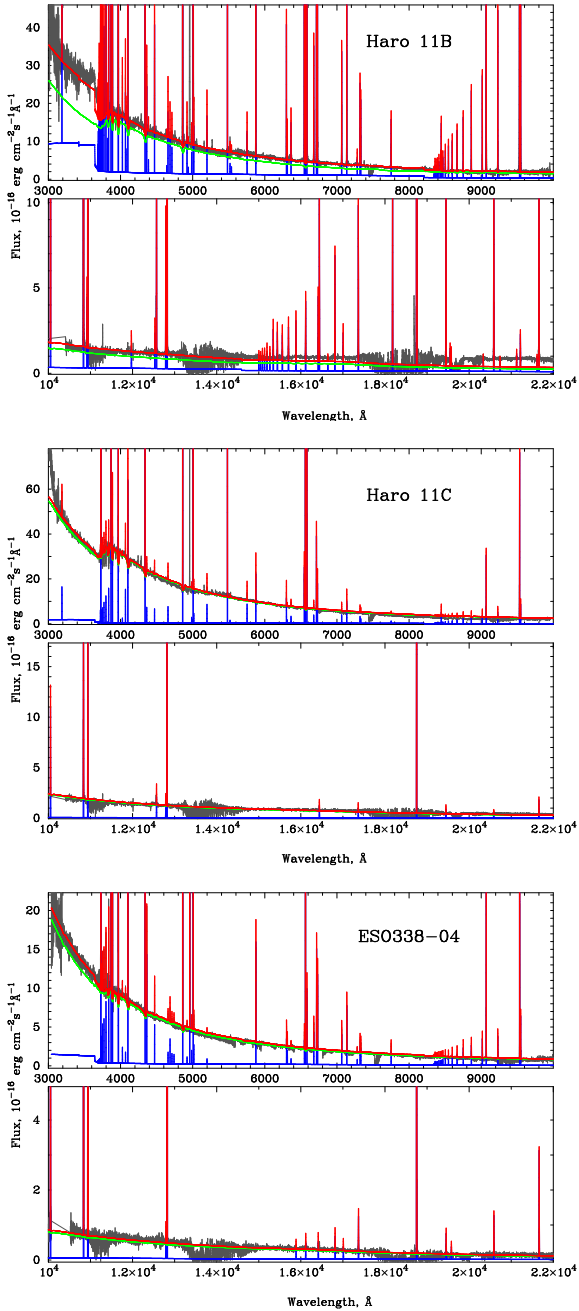

\vspace{0.3cm}
\hspace*{0.0cm}\psfig{figure=fig6a.ps,angle=-90,width=7.5cm,clip=}
\vspace{0.3cm}
\hspace{12.5cm}
\hspace*{0.0cm}\psfig{figure=fig6b.ps,angle=-90,width=7.5cm,clip=}
\vspace{0.3cm}
\hspace{12.5cm}
\hspace*{0.0cm}\psfig{figure=fig6c.ps,angle=-90,width=7.5cm,clip=}
\caption{Best-fit model SED fits to the redshift- and 
extinction-corrected observed spectra of Haro 11B, Haro 11C, and 
ESO 338-IG 004. 
The contributions from the stellar and 
ionised gas components are shown by the green and blue
lines, respectively. The sum of stellar and ionised gas emission
is shown by the red line.
}
\label{sed}
\end{figure}

The observed $F$($\lambda$) and extinction-corrected $I$($\lambda$) fluxes
of the emission lines, and the extinction coefficients $C$(H$\beta$) 
are shown in Table \ref{Int}.
   The extinction coefficients $C$(H$\beta$) = 0.74, 0.27, and 0.31
for Haro 11B, C, and ESO 338-IG 004, respectively, are derived from the 
hydrogen Balmer lines only in the UVB and VIS arm spectra. They
are adopted for the correction of emission-line fluxes not only in the
UVB and VIS spectra, but also in the NIR range.
 We also show in Table \ref{Int} the theoretical recombination fluxes 
for hydrogen Balmer, Paschen, and Brackett lines calculated by \citet{HS87} 
(last column of the Table), adopting a case B model with an electron 
temperature $T_{\rm e}$ = 12500 K, and an electron number density 
$N_{\rm e}$ = 100 cm$^{-3}$.

 The deviations of
the observed fluxes of bright Balmer, 
Paschen and Brackett hydrogen lines after correction for  
extinction with a single value of $C$(H$\beta$)
from the theoretical recombination values \citep{HS87}
do not exceed a few percent (cf. Table \ref{Int}). 
  The exceptions are extinction-corrected fluxes of several Paschen and 
Brackett hydrogen lines marked in 
Table~\ref{Int}, which deviate significantly from the theoretical values
(e.g., Pa$\beta$ line) because of strong telluric absorption.
 Excluding these lines, we may conclude that a single $C$(H$\beta$) value 
derived from the hydrogen Balmer lines can be used over the entire 
wavelength range of $\sim$$\lambda\lambda$3000 -- 24000\AA.

  Thus, we confirm our previous findings obtained for several low-metallicity 
emission-line galaxies (Mrk 59, 
II Zw 40, Mrk 71, Mrk 996, SBS 0335--052E, PHL 293B, and GRB HG 031203)
that the extinction coefficient $C$(H$\beta$) is not systematically higher 
across the NIR wavelength range than the optical range and that 
no additional regions contribute to the NIR line emission
that are, however, hidden in the optical range 
\citep{Vanzi00,Vanzi02,I09,IGFH11,IT11,Guseva2011}.

\subsection{H$_2$ ro-vibrational emission \label{s:H2}}

Molecular hydrogen lines do not originate in H {\sc ii} regions,
but in neutral molecular clouds. In the NIR, they are
excited by two mechanisms. 
  The first is a thermal mechanism
consisting of collisions between neutral species (e.g., H,
H$_2$), resulting from large-scale shocks driven by powerful stellar
winds, supernova (SN) remnants, or molecular cloud-cloud collisions.
  The second is fluorescence due to the absorption
of ultraviolet photons produced by hot stars.
  By comparing the observed line ratios with those predicted by for instance the 
models of \citet{Black1987}, 
it is possible to discriminate between the two processes. In
particular, line emission from the vibrational level $v$ = 2 and
higher vibrational levels are virtually absent in collisionally
excited spectra, while they are relatively strong in fluorescent
spectra.

  Thirteen ro-vibrational H$_2$ emission lines are detected in the NIR spectrum
of Haro 11B (see Fig.~\ref{sp_Haro11B}).
 Only two H$_2$ emission lines, 1.957$\mu$m 1-0 $S$(3) and 
2.122$\mu$m 1-0 $S$(1), are 
detected in the spectrum of ESO 338-IG 004 and no H$_2$ lines are 
detected in the spectrum of Haro 11C (Table~\ref{Int}).
  The fluxes of the  H$_2$ lines relative
to that of the strongest 2.122 $\mu$m 1-0 $S$(1) line are shown
in Table~\ref{H2}. 
  Two H$_2$ emission lines, the 1.188 $\mu$m 2-0 $S$(0) and the  
1.238 $\mu$m 2-0 $Q$(1) transitions are blended with the Fe {\sc ii} lines. 
The very large observed flux ratio of the 1.188 $\mu$m line
(Table~\ref{H2}) can be due to contamination by a singly ionised
iron line. The last two columns show the theoretical
ratios calculated by \citet{Black1987} in the
cases of fluorescent and collisional excitation.
  It can be seen that the observed line flux ratios for most of the H$_2$ lines 
 agree with those predicted for fluorescent excitation,
supporting the previous findings that fluorescence
is the main excitation mechanism of ro-vibrational H$_2$ lines in most BCDs 
with high-excitation H {\sc ii} regions \citep{Vanzi00,Vanzi08,I09a,IT11}. 
 However, there are still some BCDs \citep[e. g., PHL 293B ][]{IGFH11} for which
the collisional excitation of ro-vibrational H$_2$ lines is the dominat excitation.

\subsection{CLOUDY models \label{CLOUDY}}

  Using the CLOUDY code (version c10.00) of \citet{Ferland98}, 
we examine the excitation mechanisms
not only of the hydrogen lines  
but all emission lines by producing 
stellar photoionisation models for our objects. 
The CLOUDY input parameters 
for the best-fit models with a single stellar population
are shown in Table \ref{model}.
 The logarithm of the number of ionising photons per second $Q$(H) 
is calculated from the extinction-corrected H$\beta$ luminosity.
We adopt the shape of the ionising stellar
radiation of the Starburst-99 models \citep{Leitherer99} with a starburst
age shown in Table \ref{model}.
  The electron number density $N_{\rm e}$ is assumed to be 
constant with radius.
 The ratios of abundances of different species to hydrogen are taken from 
this paper, except for carbon which has no bright emission lines in the
optical and NIR ranges. 
  To obtain the abundance of this species, we use a
relation C/O vs. oxygen abundance for the emission-line galaxies studied 
by \citet{Garnett95}.
 The emission-line fluxes for
the best-fit models are shown in Table \ref{CLOUD}.
  
 The CLOUDY-predicted 
and extinction-corrected fluxes of bright emission lines in Haro 11B and 
Haro 11C are shown in Table \ref{CLOUD}. 
  The overall agreement achieved implies 
that a H {\sc ii} region model including only the stellar 
ionising radiation of a single stellar population is able to 
account for the observed fluxes, in both the optical and NIR ranges.
No additional excitation mechanism such as shocks from stellar 
winds and supernova remnants is needed. This conclusion also holds for the
[Fe {\sc ii}] $\lambda$12570,16436\AA\ emission lines ( Table \ref{CLOUD}),
which are often considered as shock indicators \citep[e.g. ][]{MO88,O90,O01}.

  We note that the photoionisation models for a single stellar population 
cannot reproduce the extinction-corrected fluxes in ESO 338-IG 004,
because they are measured in the spectrum that includes two
star-forming regions with different excitation properties. The
superposition of two CLOUDY models should be used instead to fit
the observations of ESO 338-IG 004.

\begin{table} [t]
\caption{H$_2$ emission line flux ratios for Haro 11B  \label{H2}}
\begin{tabular}{llccc} \hline \hline
$\lambda$& Line & Observations$^{\rm a}$ &\multicolumn{2}{c}{Model$^{\rm b}$}   \\  \cline{4-5}   
$\mu$m&& & Fluor.& Collis. \\ \hline 
1.162& H$_2$ 2-1 $S$(1)    & 0.2 & 0.4 & 0.0   \\
1.185& H$_2$ 3-1 $S$(3)    & 0.2 & 0.4$^{\rm c}$ & ...    \\ 
1.188& H$_2$ 2-0 $S$(0)+[Fe {\sc ii}] & 0.9 & 0.2$^{\rm c}$ & ...    \\
1.207& H$_2$ 3-1 $S$(2) & 0.1 & 0.2 & 0.0    \\ 
1.233& H$_2$ 3-1 $S$(1) & 0.6 & 0.5 & 0.0    \\
1.238& H$_2$ 2-0 $Q$(1)+Fe {\sc ii}? & 0.2 & 0.4 & 0.0    \\
1.601& H$_2$ 6-4 $Q$(1) & 0.1 & 0.4 & 0.0    \\ 
1.957& H$_2$ 1-0 $S$(3) & 0.5 & 0.7$^{\rm c}$ & ...    \\ 
2.034& H$_2$ 1-0 $S$(2) & 0.3 & 0.5 & 0.3    \\ 
2.073& H$_2$ 2-1 $S$(3) & 0.3 & 0.2 & 0.0    \\ 
2.122& H$_2$ 1-0 $S$(1) & 1.0 & 1.0 & 1.0    \\ 
2.223& H$_2$ 1-0 $S$(0) & 0.2 & 0.6 & 0.3    \\ 
2.248& H$_2$ 2-1 $S$(1) & 0.2 & 0.5 & 0.0    \\ \hline
\end{tabular}
\hspace{1.cm}$^{\rm a}$Flux ratios relative to the H$_2$ 2.122 $\mu$m flux. \\
\hspace{1.cm}$^{\rm b}$Model values by \citet{Black1987}. We adopt their model 1 
for fluorescent lines and model S1 for collisionally excited lines. \\
\hspace{1.cm}$^{\rm c}$Flux ratios are from model 14 by \citet{Black1987},
because these lines are not present in model 1. \\
\end{table}


\begin{table} [t]
\caption{Input parameters for the 
CLOUDY model
 \label{model}}
\begin{tabular}{lcc} \hline 
Parameter & Haro 11B & Haro 11C  \\ \hline
log $Q$(H)$^{\rm a}$         & 53.51 & 52.30  \\
Starburst age, log Myr & 6.35 & 6.50  \\
$N_{\rm e}$, cm$^{-3}$        & 10 & 10     \\
$f$$^{\rm b}$                & 0.02 & 0.06    \\
log He/H               & --0.93 & --0.93  \\
log C/H                & --3.97 & --4.20  \\
log N/H                & --4.58 & --4.69  \\
log O/H                & --3.67 & --3.90  \\
log Ne/H               & --4.33 & --4.53  \\
log S/H                & --5.50 & --5.64  \\
log Ar/H               & --6.17 & --6.22  \\ \hline
\end{tabular}

$^{\rm a}$$Q$(H) is the number of ionising photons per second. \\
$^{\rm b}$Volume filling factor. \\
\end{table}


\begin{table} [t]
\caption{Comparison of the observed and predicted intensities of some strong
emission lines in Haro 11B and C  \label{CLOUD}}
\begin{tabular}{lcccc} \hline \hline
 Ion & \multicolumn{2}{c}{  CLOUDY} &\multicolumn{2}{c}{Observations$^{\rm a}$}   \\  
\cline{2-5} 
 & B$^{\rm b}$ &  C$^{\rm b}$ & B$^{\rm b}$ &  C$^{\rm b}$ \\ \cline{2-5}   
     &$I$/$I$(H$\beta$)&$I$/$I$(H$\beta$) 
&$I$/$I$(H$\beta$)&$I$/$I$(H$\beta$) \\ \hline 
3727 [O {\sc ii}]$^{\rm c}$& 2.058& 1.927  & 2.186 & 2.336  \\
3869 [Ne {\sc iii}] & 0.283 & 0.321  & 0.299 & 0.314 \\
4101 H$\delta$      & 0.264 & 0.267  & 0.277 & 0.269    \\
4340 H$\gamma$      & 0.472 & 0.475  & 0.499 & 0.461    \\
4861 H$\beta$       & 1.000 & 1.000  & 1.000 & 1.000    \\
4959 [O {\sc iii}]  & 1.208 & 1.070  & 1.194 & 0.990     \\
5007 [O {\sc iii}]  & 3.635 & 3.220  & 3.716 & 3.172   \\
6563 H$\alpha$      & 2.917 & 2.891  & 2.907 & 2.864     \\
6583 [N {\sc ii}]   & 0.342 & 0.381  & 0.488 & 0.576   \\
6716 [S {\sc ii}]   & 0.072 & 0.078  & 0.156 & 0.291    \\
6731 [S {\sc ii}]   & 0.051 & 0.055  & 0.102 & 0.134    \\
9069 [S {\sc iii}]  & 0.172 & 0.154  & 0.156 & 0.144    \\
9532 [S {\sc iii}]  & 0.426 & 0.382  & 0.432 & 0.385    \\
10829 He {\sc i}    & 0.286 & 0.254  & 0.326 & 0.283   \\
12570 [Fe {\sc ii}] & 0.021 & 0.030  & 0.014& 0.028    \\
12821 Pa$\beta$     & 0.164 & 0.161  & 0.144 & 0.176    \\
16436 [Fe {\sc ii}] & 0.018 & 0.025  & 0.012& 0.017    \\
19450 Br$\delta$    & 0.018 & 0.018  & 0.019& 0.014    \\
21661 Br$\gamma$    & 0.028 & 0.027  & 0.031& 0.028      \\ \hline
  \multicolumn{5}{c}{MIR data for Haro 11}  \\ \hline
 Ion & $F$$^{\rm d}$&$I$/$I$(H$\beta$)$^{\rm e}$
&$I$/$I$(H$\beta$)$^{\rm f}$&CLOUDY \\ \hline
10.51$\mu$m [S {\sc iv}]   &4.395 & 0.122 & 0.365 & 0.224 \\         
12.81$\mu$m [Ne {\sc ii}]  &3.123 & 0.087 & 0.259 & 0.097 \\           
15.55$\mu$m [Ne {\sc iii}] &9.814 & 0.273 & 0.814 & 0.466 \\ 
18.71$\mu$m [S {\sc iii}]  &4.524 & 0.126 & 0.376 & 0.206 \\ \hline
\end{tabular}
\hspace{1.cm}$^{\rm a}$extinction-corrected fluxes. \\
\hspace{1.cm}$^{\rm b}$B and C mean Haro 11B and Haro 11C, 
respectively. \\
\hspace{1.cm}$^{\rm c}$sum of the 3726 and 3728 [O {\sc ii}] lines. \\
\hspace{1.cm}$^{\rm d}$observed flux in units of 10$^{-13}$ erg s$^{-1}$ cm$^{-2}$
\citep{Wu2008}. \\
\hspace{1.cm}$^{\rm e}$extinction-corrected flux ratios with $C$(H$\beta$)=0.745. \\
\hspace{1.cm}$^{\rm f}$extinction-corrected flux ratios with $C$(H$\beta$)=0.27. \\
\end{table}


  To extend our study of Haro 11 into the MIR range, we 
used the {\sl Spitzer} observations of \citet{Wu2008} obtained with 
an aperture, which is much larger than 
the one used in the X-shooter observations,
and includes all of the bright H {\sc ii} regions A, B, and C. Therefore, to
compare the observed and CLOUDY-predicted emission-line fluxes 
in the MIR
range (relative to H$\beta$) we should take into account aperture corrections. 
In particular, the extinction-corrected H$\beta$ flux from the X-shooter 
observations cannot be used because it was obtained with a smaller aperture.
 We instead derive the H$\beta$ extinction-corrected flux from the total 
H$\alpha$ flux $F$(H$\alpha$)=3.326$\times$10$^{-12}$ erg s$^{-1}$ cm$^{-2}$ 
measured by \citet{Wu2008} after its correction for extinction.
We do not know the average extinction coefficient $C$(H$\beta$) 
within an aperture used for the MIR observations by \citet{Wu2008}.
 It is known from {\sl HST} data \citep{Adamo2010} that
the reddening in Haro 11 varies in the range
$E$($B-V$)=0.0 to 0.6 (corresponding to $C$(H$\beta$)=0.0 to 0.88),
and to even higher values, in agreement with our extinction determinations
for Haro 11B and Haro 11C. 
  Therefore, to correct the H$\beta$ flux for
extinction within a large aperture, we consider two limiting cases 
by adopting the $C$(H$\beta$) values derived from both the Haro 11B spectrum (high
extinction) and the Haro 11C spectrum (low extinction).
It is also reasonable to assume that the averaged extinction in the
area covered by the MIR observations is between these two limiting
cases.
  In the third and fourth columns of Table~\ref{CLOUD}, we show the MIR 
emission-line fluxes normalised to the extinction-corrected H$\beta$
flux with 
$C$(H$\beta$)=0.745 (the case of Haro 11B) and $C$(H$\beta$)=0.27 
(the case of Haro 11C), respectively. 
We also note that a correction of
MIR emission lines for extinction is not needed.
 In the last column of Table~\ref{CLOUD} for MIR data, we show 
CLOUDY predictions 
with input parameters from Table \ref{model}.
 It is seen from Table \ref{CLOUD} that the predicted relative MIR emission-line
fluxes are between the two limiting cases of extinction-corrected fluxes
with high and low extinctions, suggesting that with a more realistic 
intermediate extinction the agreement between models and observations
would be much better. 
  This implies that the MIR range, similar to the NIR range,
does not resemble highly extincted emission-line regions that significantly
contribute to the MIR line emission.

\subsection{Evidence for luminous blue variable (LBV) stars in Haro 11 \label{LBV}}

The H$\alpha$ emission line in the spectrum of Haro 11C is asymmetric
and has also a broad component.
\citet{OstlinAmram2001} also reported that the H$\alpha$ line profiles 
of Haro 11 are broad, up to full widths at half maximum ($FWHM$) 
of 270 km s$^{−1}$, and have a non-Gaussian shape.
They suggested that two or more non-virialised ionised gas components
may be present in the galaxy.
  The extinction-corrected relative H$\alpha$ flux
$I$(H$\alpha$)/$I$(H$\beta$) = 3.82 in Haro 11C, obtained with 
$C$(H$\beta$)=0.27, deviates significantly from the theoretical 
recombination value 2.86 (\citet{HS87} for the electron temperature 10000 K 
and the electron number density $N_{\rm e}$ = 100 cm$^{-3}$ when the observed 
H$\alpha$ flux $F$(H$\alpha$)=3.350$\times$10$^{-14}$ erg s$^{-1}$ cm$^{-2}$
is adopted. This high H$\alpha$ flux may be due to substantial 
emission from either supernovae of type II or 
circumstellar envelopes of stars.
  Because of its anomalous behavior this line was excluded in sect. 
\ref{abund} from the calculation of the extinction coefficient.
The decomposition of the H$\alpha$ emission line of Haro 11C into 
three  
Gaussian emission components and one Gaussian absorption component is shown 
in Fig.~\ref{d_Ha_2f_NII}. 
   Two of the Gaussian emission components (the narrowest and the broadest) 
marked by blue 
dotted lines are likely linked to the nebular emission, implying that there is
low- and high-velocity ionised gas in the H {\sc ii} region. 
These two components are sufficient to reproduce the shape of the strongest
forbidden line [O {\sc iii}] $\lambda$5007, strengthening the evidence for a 
nebular origin of these components. 
   However, the H$\alpha$ profile is more complex,
and two additional components (emission and absorption) are needed.
The green dotted line is the composition of the third emission and
absorption components and represents a P Cyg profile.
The observed H$\alpha$ profile (black solid line)
is nicely fit by the summed profile, consisting of two nebular components
and the P Cyg profile (red dashed line).
 The flux $F$(H$\alpha$)=2.281$\times$10$^{-14}$ erg s$^{-1}$ cm$^{-2}$
of the H$\alpha$ line after subtraction of the P Cyg profile corresponds
to the recombination ratio 
$I$(H$\alpha$)/$I$(H$\beta$) = 2.86 (Table~\ref{Int}) after the correction
for extinction with $C$(H$\beta$) = 0.27 is applied.
  Thus, the inclusion of the stellar component of H$\alpha$ emission is required
to fit the observations, implying that either LBV star(s) or 
supernovae are present. 
 The H$\alpha$ emission line is therefore the most affected by the presence 
of a stellar wind. 
The observed Balmer decrement of the broad emission lines
in LBVs and SNe is often much larger than that for the narrow lines,
reflecting the high density and possibly high extinction in the circumstellar 
envelopes where the broad lines
originate. In particular, this is true for the LBV in PHL 293B and results 
in a very strong H$\alpha$ emission line compared to the H$\beta$ line
\citep{IT09,IGFH11}. On the other hand, the small $FWHM$
of $\sim$ 180 km s$^{-1}$ of the P Cyg component precludes the supernovae scenario.
We note that, in principle, in ESO 338-IG 004 some additional 
contribution from LBV stars or shocks to the H$\alpha$ emission can also 
be present. 
Therefore, $C$(H$\beta$) was derived excluding H$\alpha$,  
hence $I$(H$\alpha$)/$I$(H$\beta$) = 2.94 
(Table~\ref{Int}) is slightly higher than the recombination ratio.

\begin{table} [t]
\caption{Eqiuvalent widths of the stellar absorption CO band 
($\lambda$=2.3$\mu$m) in Haro 11  \label{CO}}
\begin{tabular}{llccccc} \hline \hline
 &&& \multicolumn{4}{c}{Starburst-99$^{\rm a,b}$}   \\  
\cline{4-7}   
 & &&\multicolumn{2}{c}{inst.$^{\rm c}$} & \multicolumn{2}{c}{cont.$^{\rm d}$} \\ 
\cline{4-7} 
\multicolumn{2}{c}{Observations}&& $Z$=0.004 & $Z$=0.008 & $Z$=0.004 &$Z$=0.008 \\ \cline{1-2} \cline{4-7}
Haro 11B& 8\AA\  && 8-11\AA\ & 13-15\AA\ & 2-8\AA\ & 8-10\AA\ \\
Haro 11C& 9\AA\  &&  \\ \hline
\end{tabular}
\smallskip

$^{\rm a}$Values for the time of 8--10 Myr after the onset of star formation 
\citep{Leitherer99}. \\
$^{\rm b}$$Z$ is heavy element mass fraction. \\
$^{\rm c}$Instantaneous burst model. \\
$^{\rm d}$Continuous star formation.
\end{table}


The extinction-corrected luminosity of the H$\alpha$ P Cyg 
component in Haro 11C is very high, $L$(P Cyg) = 1.32$\times$10$^{40}$ 
erg s$^{-1}$, exceeding by a factor of $\sim$ 10 -- 100 the H$\alpha$ 
luminosities of LBV stars in the low-metallicity emission-line 
galaxies NGC 2366 
\citep{D97,D01}, PHL 293B, and DDO 68 \citep{I07,IT09,IGFH11}.

  Moreover, sixteen permitted Fe {\sc ii} emission
lines are observed in the spectra of Haro 11B and Haro 11C  
(see Figs.~\ref{sp_Haro_11C_detail}, \ref{sp_Haro11B}, \ref{sp_Haro_11C},
\ref{sp_E338} and Table~\ref{Int}), which reflects their circumstellar origin
in the dense stellar wind. No such lines are present in low-density 
nebular gas. The Orion Nebula serves 
as a template for the investigation of different excitation mechanisms of 
emission  
lines observed in a gaseous nebulae. It is one of the most well-studied of nearby 
Galactic H {\sc ii} regions with an extensive list of identifications 
of various lines, including the very weak ones 
\citep[e.g.][]{Peimbert1977,Osterbrock1992}. There are 
high-resolution spectroscopic observations of the Orion nebula with high
signal-to-noise ratios. Nevertheless, the Fe {\sc ii} lines observed in the
VLT/X-shooter spectra of Haro 11B and C are not present in the spectra of 
the Orion nebula obtained by \citet{Baldwin2000},  \citet{Esteban2004}, and 
\citet{Mesa2009}. 

   On the other hand, the permitted Fe {\sc ii} emission lines 
are common in spectra of LBV stars 
undergoing giant eruption events during the late evolutionary stages 
of the most massive 
stars. One of the most well-known LBVs is the binary 
system $\eta$ Carina ($\eta$ Car), which undergoes periodical outbursts
\citep[e.g.][]{Massey2003,Smith2004,Damineli2008,Clark2005,Damineli2008a,Soker2011}.
  The bright emission lines of H, He, and the permitted 
Fe {\sc ii} and forbidden [Fe {\sc ii}] lines 
are observed in LBVs that also display 
P Cyg profiles during eruption events. 
  Most of the emission lines of different 
species (Mg {\sc i}, [Fe {\sc ii}], Si {\sc ii}, [Ni {\sc ii}], [Cr {\sc ii}],
N {\sc i}, and [C {\sc i}]) (Table~\ref{Int}) are observed in nebulae as well as 
in emission-line stars like LBVs, supergiants with H and He emission, and 
Of/WN9 stars. 
We also note that all Fe {\sc ii} emission lines seen in the spectra of
Haro 11B and Haro 11C are detected only in the high-resolution spectra
of $\eta$ Car by e.g. \citet{Hamann1994} and \citet{Wallerstein2001}.
For the line identification, we use the data for $\eta$ Car 
of \citet{Wallerstein2001} in a wide wavelength range  
$\sim$$\lambda$3100 -- 9000\AA\ (with several gaps due to observational program
reasons or telluric absorption) and data of \citet{Hamann1994} in 
the wavelength range $\sim$$\lambda$6450 -- 24500\AA. These two sets of data
cover the whole wavelength range of the VLT/X-shooter observations excluding
gaps. Additionally, for some line identifications 
we used the data of \citet{Gaviola1953}, \citet{Johansson1977}, and
\citet{Allen1985}. 

We find that LBV is probably also present in Haro 11B. This is indicated by the
Fe {\sc ii} and [Fe {\sc ii}] emission lines in its spectrum 
(Fig.~\ref{sp_Haro_11C_detail}, Table~\ref{Int}). It is also possible 
that broad H$\alpha$
emission is also present, but masked by very strong nebular
H$\alpha$ and [N {\sc ii}] $\lambda$6548, 6583\AA\ emission lines.
On the other hand, no permitted  Fe {\sc ii} emission lines were detected 
in the spectrum of ESO 338-IG 004 (see Table~\ref{Int}). 

  Thus, we have three pieces of evidence supporting the idea that a LBV star is
present in Haro 11C. These are that  a) a P Cyg profile is needed to reproduce
the observed H$\alpha$ profile, b) the only possible way to
adjust the observed Balmer decrement 
to the theoretical one is the subtraction of the P Cyg component 
from the H$\alpha$ 
emission line, and c) the presence of numerous Fe {\sc ii} emission lines, 
which is a common feature of the LBV star spectrum.

 \citet{Smith2011} collected data for a number of 
extragalactic optical transients
or ``supernova impostors'' related to giant eruptions of LBVs and illustrated 
their remarkably wide diversity in terms of their peak absolute
magnitude, duration, progenitor stars, outburst spectra, and 
other observable properties. In particular, the peak absolute magnitude varies 
from --10 mag to --16 mag and increases during eruption by more than 2 -- 4 mag 
for LBV stars of the Milky Way 
($\eta$ Car, P Cyg), V1 in NGC 2366, HD 5980 in the SMC, and others. 
There is not much information in the literature about the range of 
LBV H$\alpha$ luminosities, but it
is likely that they have a large range,
which is similar to the absolute magnitudes.
Therefore, it is difficult to quantitatively estimate the number of LBV stars 
in Haro 11C.

\subsection{Spectral energy distributions, stellar masses and a red continuum 
excess \label{SED}}

   We derive the stellar masses of three H {\sc ii} regions 
(see Table~\ref{comp}) by fitting 
their spectra to a superposition of SEDs of
single stellar populations with different ages following
the prescriptions of \citet{G2006,G2007} and \citet{IGT2011}. 
 This technique also 
allows us to check whether a red continuum excess is present
in the spectra of Haro 11B and Haro 11C.
  We note, that for star-forming galaxies with  EW(H$\beta$) $\ge$ 100\AA\
the ionised gas continuum emission contributes appreciably to the total 
continuum emission. 
Neglecting it can lead to an overestimation of 
galactic stellar masses by a factor of a few \citep{IGT2011}.
  We include the ionised gas emission in the fit 
of the spectra of all three objects, despite the equivalent width
of the H$\beta$ emission line being large (EW(H$\beta$)=155\AA) only in the 
Haro 11B spectrum, while EW(H$\beta$) $<$ 100\AA\ in the spectra of Haro 11C and 
ESO 338-IG 004 (Table~\ref{Int}).

As each SED is the sum of both stellar and ionised gas
emission, its shape depends on the relative contributions of these 
two components. In active star-forming regions, the contribution of the  
ionised gas emission can be very large. However,  
the equivalent widths of the H$\beta$ 
emission line never reach the theoretical values for
pure gaseous emission of $\sim$ 900 -- 1100\AA, implying that there is a non-zero 
contribution of stellar emission in all objects. To take into account
gaseous emission, we use the observed emission-line fluxes and equivalent
widths. The contribution of the gaseous emission is scaled 
to the stellar emission using the ratio of the observed equivalent width of 
the H$\beta$ emission line to the equivalent width of H$\beta$ expected 
for pure gaseous emission. 
The continuous gaseous emission is taken from \citet{Aller1984} and 
includes hydrogen and helium free-bound, free-free and 
two-photon emission. In our models, this emission is always calculated using 
the electron temperature $T_e$(H$^+$) of the H$^+$ zone and with the He/H 
abundance ratio derived from the H {\sc ii} region spectrum.

    To calculate the contribution of stellar emission to the SEDs,
we adopt a grid of the Padua stellar evolution models of 
\citet{Girardi2000}\footnote{http://pleiadi.pd.astro.it.} 
with a heavy element mass fraction $Z$ = 0.008 for Haro 11B and $Z$ = 0.004
for Haro 11C and ESO 338-IG 004. Using these data, we calculate with the
package PEGASE.2 \citep{FR97} a grid of stellar
instantaneous burst SEDs for a range of ages from  0.5 Myr to 15 Gyr.
We adopt a stellar initial mass function with a Salpeter slope, an
upper mass limit of 100 $M_\odot$, and a lower mass limit of 0.1 $M_\odot$.
 The SED for any star formation history (SFH) can then be obtained by 
integrating the instantaneous burst SEDs over time with a specified 
time-varying star formation rate.
  We approximate the SFH by a combination of a recent short burst to 
account for the young stellar population,
and continuous star formation to describe the properties of the older stars.

  Finally, the contribution of the nebular continuum and emission lines 
were taken into account. Each fit was performed over the whole observed 
spectral range $\sim$$\lambda$$\lambda$3000 -- 24000\AA. 
  We performed 5$\times$10$^5$  Monte Carlo models by
varying simultaneously the age of young stellar population $t$(young), 
the age of the oldest stars $t$(old), 
the mass ratio of old-to-young stellar populations
 $b$ = $M$(old)/$M$(young), 
the electron temperature $T_{\rm e}$(H$^+$) in the H$^+$ zone, and the 
extinction coefficients for gas and for stars,
$C$(H$\beta$)$_{\rm gas}$ and $C$(H$\beta$)$_{\rm stars}$, respectively.
   
  The best-fit model SED is found from $\chi ^2$ minimisation of the
deviation between the modelled and the observed continuum in ten ranges 
of the whole spectrum free of emission lines.

    In Fig.~\ref{sed}, we show the best-fit model SED fits to the redshift- and 
extinction-corrected spectra of Haro 11B, C, and ESO 338-IG 004. 
   The contributions from the stellar and 
ionised gas components are shown by green and blue
lines, respectively. 
The sum of both stellar and ionised gas emission is shown by a red line.
   It is seen that the contribution of gaseous emission is essential 
only for Haro 11B because of the high EW(H$\beta$). 
For an other two objects, this contribution is small,
especially for Haro 11C with EW(H$\beta$) = 16\AA. 
 
  \citet{Adamo2010} found using {\sl HST} photometry data for Haro 11,  
that a strong red excess is present in spectra
of Haro 11B and Haro 11C at wavelengths $>$ 8000\AA\ with respect to the 
synthetic evolutionary models.
 They show that despite the tighter model fit to the Haro 11B spectrum
achieved ($\chi ^2$ for Haro 11B is smaller than that for Haro 11C), the red excess 
in Haro 11B is larger. 
Fig.~\ref{sed} shows that all observed spectra in the whole 
range of wavelengths 
$\sim$$\lambda$$\lambda$3000--24000\AA\ are fitted 
quite well by the model SEDs, 
except for the spectrum of Haro 11B where the flux excess at 
wavelengths $>$1.6$\mu$m is clearly present. 
 Thus, \citet{Adamo2010} found red excesses in the B and C knots but in 
the present paper we find one only for the B knot. This disagreement can be 
partially associated with the different sizes of the extracted regions 
(we adopted extraction apertures of $\sim$1\arcsec $\times$ 4\arcsec\ for 
Haro 11B and C, while \citet{Adamo2010} adopted much smaller apertures).
However, we note that the majority of the light in the B and C knots
within the large adopted spectroscopic apertures comes from the brightest 
compact clusters within the slit, implying that we measure similar continuum 
flux densities to those in the photometric study of \citet{Adamo2010}.

What is the nature of this excess?
 As mentioned above the observed fluxes of the Balmer, 
Paschen, and Brackett hydrogen lines after correction for  
extinction with a single value $C$(H$\beta$) 
(see Table \ref{Int}) agree well with the 
theoretical recombination values of \citet{HS87} for all studied 
star-forming regions, i.e. Haro 11B, Haro 11C, and ESO 338-IG 004. 
 The same result was obtained from CLOUDY modelling (Table~\ref{CLOUD}). 
   This implies that the NIR excess 
is not due to highly obscured hot stars ionising the interstellar
medium, which would require that the continuum excess be followed by an excess 
of the NIR emission lines, a result we do not find.
Evidently, only the non-ionising cool stars, including a large
population of luminous red supergiant 
stars (RSGs), can contribute to the NIR continuum excess.

This conclusion is supported by the detection of the strongest stellar 
absorption CO band at $\sim$2.3 $\mu$m in the spectra of Haro 11B and Haro 11C.
This spectral feature is a signature of 
red giant and red supergiant stars. The equivalent width 
of the CO band depends on 
stellar temperature and luminosity and decreases with both increasing 
stellar temperature and decreasing stellar luminosity. 
We measure equivalent widths EW(CO 2.3 $\mu$m) of 8\AA\ and 9\AA\ in Haro 11B
and Haro 11C, respectively. In Table~\ref{CO}, we compare the observed values 
with the Starburst-99 
stellar population synthesis models \citep{Leitherer99} and
conclude that the stellar population of an older burst with an
age of $\sim$ 10 Myr including red supergiant stars 
can be responsible for the observed red excess
and CO absorption at $\sim$2.3 $\mu$m. This older
non-ionising stellar population is probably more obscured by the dust and therefore 
does not contribute to the optical light. However, we are unable to infer whether
this is true from the hydrogen-line decrement 
because the older population does not produce emission in hydrogen
lines.

  We derive the instantaneous $SFR$(H$\alpha$), 
using the extinction-corrected luminosity $L$(H$\alpha$) of the 
H$\alpha$ emission line and the frequently used relation by 
\citet{Kennicutt1998}, which, however, is obtained for longer-duration
star formation in disc galaxies and may not be valid for knots
B and C where stars were formed almost instantaneously.
We obtain $SFR$ = 3.5, 0.22, and 0.11 $M_{\odot}$ yr$^{-1}$ in
Haro 11B, Haro 11C, and ESO 338-IG 004, respectively, for the regions
within the apertures used for the spectroscopic observations 
(see sect.~\ref{obs}). To achieve this result, we adopt
the distances of 82.4 Mpc and 38.4 Mpc for Haro 11 and ESO 338-IG 004, 
respectively, which are taken from the NED.
The star formation rates of luminous compact 
emission-line galaxies (LCGs) derived by \citet{IGT2011} are in 
the range 0.7 - 60 $M_\odot$ yr$^{-1}$ with an average value 
of $\sim$4 $M_\odot$ yr$^{-1}$. 
Thus, Haro 11 is among the galaxies 
with significant star-formation activity in its H {\sc ii} region B.
The specific star formation rates ($SSFR$) are then defined as 
$SSFR$(H$\alpha$) = $SFR$(H$\alpha$)/$M_*$ and are equal to 
2.8$\times$10$^{-8}$ yr$^{-1}$, 1.1$\times$10$^{-9}$ yr$^{-1}$, and 
2.0$\times$10$^{-9}$ yr$^{-1}$ for Haro 11B, Haro 11C, and ESO 338-IG 004, 
respectively.
 In addition, the $SSFR$ for Haro 11B is high and similar to those found in 
LCGs by \citet{IGT2011} and in high-redshift galaxies, while the $SSFR$s 
for Haro 11C and ESO 338-IG 004 are lower by a factor of $\sim$ 10.

\section{Conclusions \label{concl}}

We have studied the spectra of the blue compact galaxies Haro 11 and 
ESO 338-IG 004 obtained by performing   
VLT/X-shooter spectroscopic observations in the 
wavelength range $\sim$$\lambda$$\lambda$3000 -- 24000\AA. 
 We have arrived at the following conclusions:

1. We derived oxygen abundances 
of 12 + log O/H = 8.33 $\pm$ 0.01 and 8.10 $\pm$ 0.04
in the two H {{\sc ii}} regions Haro 11B and Haro 11C, respectively,
which are appreciably higher than the previous determination of 
\citet {BergvallOstlin2002} (12 + log O/H = 7.9).
Our oxygen abundance 12 + log O/H = 7.89 $\pm$ 0.01 
for ESO 338-IG 004 is slightly lower than the previous determinations, which 
are in the range 7.92 -- 8.08 
\citep{Masegosa1994,Bergval1985,Raimann2000,BergvallOstlin2002,GISFHP2011}.

2. We used X-shooter data together with
{{\sl Spitzer}} observations in the mid-infrared range to test for
hidden star formation. The observed fluxes of hydrogen lines 
correspond to the theoretical recombination values after correction for  
extinction with a single value of $C$(H$\beta$) applied to a wide 
wavelength range from near-UV to NIR and MIR.
  Thus, we confirm our previous findings obtained for several low-metallicity 
emission-line galaxies (Mrk 59, II Zw 40, Mrk 71, Mrk 996, SBS 0335--052E, 
PHL 293B, and GRB HG 031203)
that the extinction coefficient $C$(H$\beta$) is not systematically higher 
in the NIR wavelength range than the optical one and that no hidden star 
formation contributes appreciably to the NIR line emission
but not the optical line emission.

3.  All observed spectra are fitted 
quite well across the entire wavelength range covered by model SEDs, 
except for the continuum flux excess at wavelengths $>$1.6$\mu$m for Haro 11B.
An anomalously large population of luminous red supergiant
stars (RSGs) in Haro 11B is probably responsible for the observed flux excess 
relative to the model SED in the NIR range.

4.  Thirteen ro-vibrational H$_2$ emission lines are detected in the 
NIR spectrum of Haro 11B. From the observed flux ratios of the H$_2$ lines, 
we conclude that fluorescence
is the main excitation mechanism of ro-vibrational H$_2$ lines. This confirms 
the previous finding for BCGs with high-excitation H {\sc ii} regions.

5. The agreement between the CLOUDY-predicted 
and extinction-corrected emission-line fluxes implies 
that a H {\sc ii} region model including only stellar 
photoionisation as an ionising source is able to account for the observed 
fluxes in the optical, NIR, and MIR ranges.
No additional excitation mechanism such as shocks from stellar 
winds and supernova remnants is needed.

6. We find evidence of an LBV star in Haro 11C:
a) a P Cyg profile is needed to reproduce the asymmetric H$\alpha$
emission line; b) the same P Cyg emission is needed to explain a strong
H$\alpha$ emission excess above the recombination value; 
c) numerous permitted singly ionised iron
emission lines are detected in the spectrum of Haro 11C.
 These lines are common in the spectra of LBV stars. The high luminosity
of the P Cyg H$\alpha$ line $L$(P Cyg) = 1.3$\times$10$^{40}$ erg s$^{-1}$
exceeds by a factor of 10 -- 100 the LBV luminosities
in some low-metallicity
emission-line galaxies.
  Similarly, the numerous Fe {\sc ii} and [Fe {\sc ii}] emission
lines in the spectrum of Haro 11B indicates that LBV is also present in this
knot. We find no evidence of LBV in ESO 338-IG 004.

\acknowledgements
N.G.G., Y.I.I., and K.J.F. are grateful to the staff of the Max Planck 
Institute for Radioastronomy for their warm hospitality. 
This research has made use of the NASA/IPAC Extragalactic Database (NED), 
which is operated by the Jet Propulsion 
Laboratory, California Institute of Technology, under contract with the 
National Aeronautics and Space Administration.

\Online

\setcounter{figure}{0}
\begin{figure*}[t]
\vspace{0.3cm}
\hspace*{1.0cm}\psfig{figure=fig1a.ps,angle=-90,width=15.0cm,clip=}
\hspace*{1.0cm}\psfig{figure=fig1b.ps,angle=-90,width=15.5cm,clip=}
\caption{Redshift-corrected VLT/X-shooter 
spectrum of Haro 11B  
with identifications of strong lines. Dubious identifications are marked 
with '?' 
}
\label{sp_Haro11B}
\end{figure*}

\begin{figure*}[t]
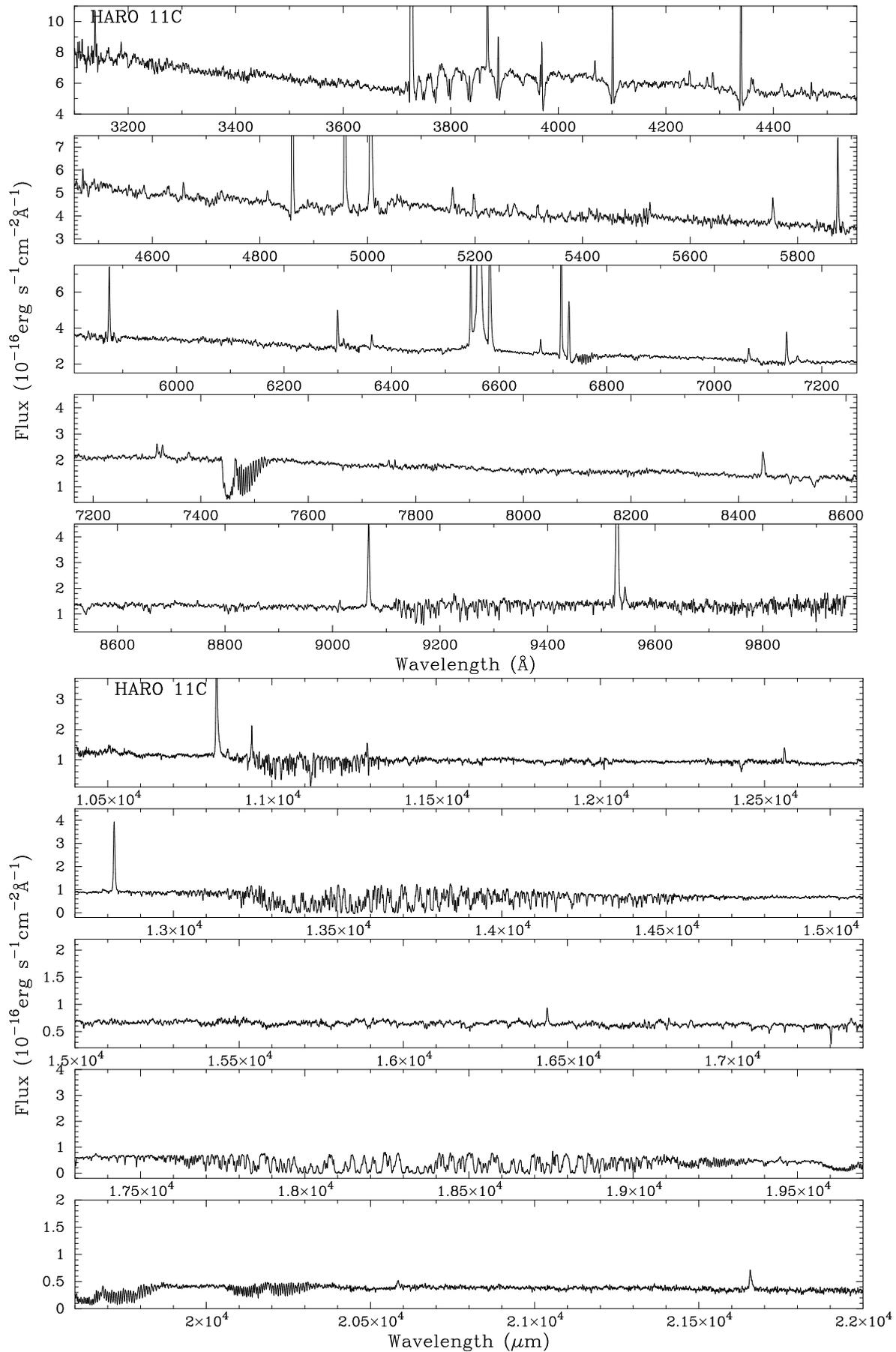

\vspace{0.3cm}
\hspace*{1.0cm}\psfig{figure=fig2a.ps,angle=-90,width=15.0cm,clip=}
\hspace*{1.0cm}\psfig{figure=fig2b.ps,angle=-90,width=15.5cm,clip=}
\caption{Same as Fig.~\ref{sp_Haro11B} but for Haro 11C.
}
\label{sp_Haro_11C}
\end{figure*}

\begin{figure*}[t]
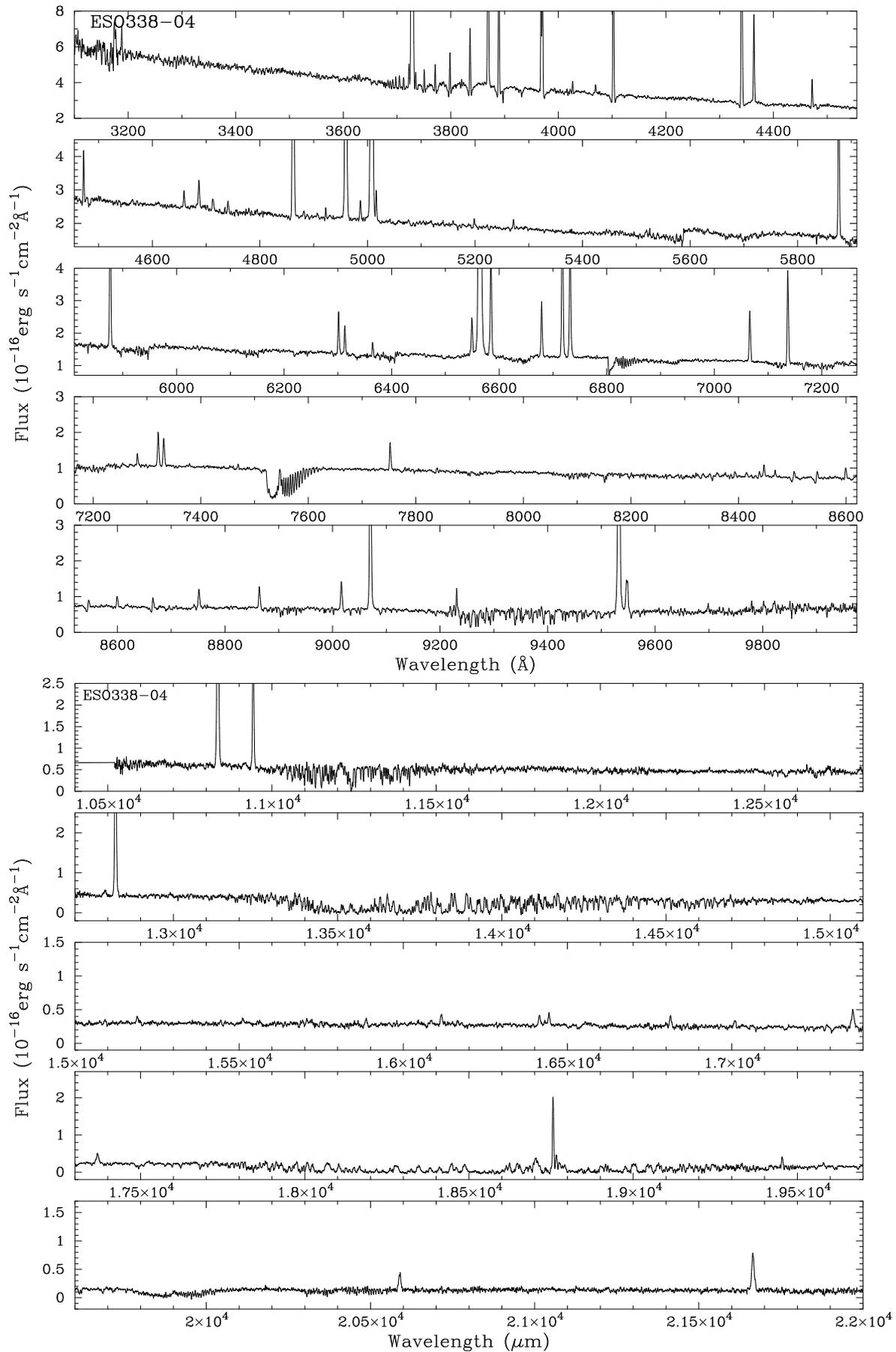

\vspace{0.3cm}
\hspace*{1.0cm}\psfig{figure=fig3a.ps,angle=-90,width=15.0cm,clip=}
\hspace*{1.0cm}\psfig{figure=fig3b.ps,angle=-90,width=15.5cm,clip=}
\caption{Same as Fig.~\ref{sp_Haro11B} but for ESO 338-IG 004.
}
\label{sp_E338}
\end{figure*}

\setcounter{table}{0}

\renewcommand{\baselinestretch}{1.0}

  \begin{longtable}{lrrcrrcrrl}
  \caption{Emission Line Intensities}
\label{Int} \\
\hline \hline
  {}&\multicolumn{8}{c}{\sc Galaxy}&{} \\          
\cline{2-9} 
  {}&
 \multicolumn{2}{c}{  Haro 11B            }&&
 \multicolumn{2}{c}{  Haro 11C            }&&
 \multicolumn{2}{c}{  ESO 338-IG 004           }&{theory}  \\ 
  \cline{2-3} \cline{5-6} \cline{8-9}                   
  {Ion}
  &{$F$($\lambda$)/$F$(H$\beta$)}$^{\rm a}$
  &{$I$($\lambda$)/$I$(H$\beta$)}$^{\rm b}$&
  &{$F$($\lambda$)/$F$(H$\beta$)}$^{\rm a}$
  &{$I$($\lambda$)/$I$(H$\beta$)}$^{\rm b}$&
  &{$F$($\lambda$)/$F$(H$\beta$)}$^{\rm a}$
  &{$I$($\lambda$)/$I$(H$\beta$)}$^{\rm b}$& {Case B$^{\rm c}$} \\ \hline \\
  {}&\multicolumn{8}{c}{a) Near-UV and optical range}\\ \\
3188 He {\sc i}                   &   2.14 $\pm$   0.07 &   4.92 $\pm$   0.16 & &   3.01 $\pm$   0.32 &   3.99 $\pm$   0.43 & &   ...~~~~~~ &   ...~~~~~~ &   ...~~~~~~ \\
3322 [Fe {\sc iii}]?              &   0.21 $\pm$   0.04 &   0.45 $\pm$   0.10 & &   ...~~~~~~ &   ...~~~~~~ & &   ...~~~~~~ &   ...~~~~~~ &   ...~~~~~~ \\
3614 He {\sc i}                   &   0.19 $\pm$   0.03 &   0.34 $\pm$   0.05 & &   ...~~~~~~ &   ...~~~~~~ & &   ...~~~~~~ &   ...~~~~~~ &   ...~~~~~~ \\
3634 He {\sc i}                   &   0.15 $\pm$   0.03 &   0.26 $\pm$   0.05 & &   ...~~~~~~ &   ...~~~~~~ & &   ...~~~~~~ &   ...~~~~~~ &   ...~~~~~~ \\
3671 H24                          &   0.16 $\pm$   0.04 &   0.31 $\pm$   0.08 & &   ...~~~~~~ &   ...~~~~~~ & &   0.19 $\pm$   0.04 &   0.24 $\pm$   0.05 & 0.39 \\
3676 H22                          &   0.24 $\pm$   0.04 &   0.40 $\pm$   0.07 & &   ...~~~~~~ &   ...~~~~~~ & &   0.42 $\pm$   0.05 &   0.53 $\pm$   0.06 & 0.50 \\
3679 H21                          &   0.25 $\pm$   0.03 &   0.43 $\pm$   0.05 & &   ...~~~~~~ &   ...~~~~~~ & &   0.34 $\pm$   0.04 &   0.43 $\pm$   0.06 & 0.57 \\
3683 H20                          &   0.30 $\pm$   0.02 &   0.51 $\pm$   0.03 & &   ...~~~~~~ &   ...~~~~~~ & &   ...~~~~~~ &   ...~~~~~~ & 0.66 \\
3687 H19                          &   0.34 $\pm$   0.02 &   0.59 $\pm$   0.03 & &   ...~~~~~~ &   ...~~~~~~ & &   ...~~~~~~ &   ...~~~~~~ & 0.77 \\
3692 H18                          &   0.49 $\pm$   0.02 &   0.85 $\pm$   0.03 & &   ...~~~~~~ &   ...~~~~~~ & &   ...~~~~~~ &   ...~~~~~~ & 0.91 \\
3697 H17                          &   0.61 $\pm$   0.02 &   1.05 $\pm$   0.04 & &   ...~~~~~~ &   ...~~~~~~ & &   ...~~~~~~ &   ...~~~~~~ & 1.08 \\
3704 H16                          &   1.06 $\pm$   0.03 &   1.82 $\pm$   0.05 & &   ...~~~~~~ &   ...~~~~~~ & &   ...~~~~~~ &   ...~~~~~~ & 1.29 \\
3712 H15                          &   0.97 $\pm$   0.03 &   1.66 $\pm$   0.05 & &   ...~~~~~~ &   ...~~~~~~ & &   ...~~~~~~ &   ...~~~~~~ & 1.57 \\
3722 H14                          &   1.82 $\pm$   0.04 &   3.09 $\pm$   0.07 & &   1.41 $\pm$   0.14 &   1.67 $\pm$   0.17 & &   ...~~~~~~ &   ...~~~~~~ & 1.93 \\
3726 [O {\sc ii}]                 &  63.69 $\pm$   0.91 & 107.80 $\pm$   1.68 & &  85.01 $\pm$   1.25 & 100.74 $\pm$   1.64 & &  46.53 $\pm$   0.70 &  57.82 $\pm$   0.94 &   ...~~~~~~ \\
3728 [O {\sc ii}]                 &  65.57 $\pm$   0.93 & 110.83 $\pm$   1.73 & & 112.19 $\pm$   1.65 & 132.86 $\pm$   2.16 & &  64.12 $\pm$   0.95 &  79.64 $\pm$   1.28 &   ...~~~~~~\\
3733 He {\sc i}                   &   2.21 $\pm$   0.13 &   3.72 $\pm$   0.23 & &   ...~~~~~~ &   ...~~~~~~ & &   ...~~~~~~ &   ...~~~~~~ &   ...~~~~~~\\
3734 H13                          &   1.62 $\pm$   0.10 &   2.73 $\pm$   0.17 & &   1.41 $\pm$   0.27 &   1.66 $\pm$   0.33 & &   1.39 $\pm$   0.07 &   1.72 $\pm$   0.09 & 2.41 \\
3750 H12                          &   1.72 $\pm$   0.05 &   3.35 $\pm$   0.13 & &   1.56 $\pm$   0.24 &   4.95 $\pm$   0.99 & &   2.28 $\pm$   0.12 &   3.35 $\pm$   0.24 & 3.07 \\
3771 H11                          &   2.20 $\pm$   0.06 &   4.12 $\pm$   0.14 & &   2.29 $\pm$   0.34 &   5.87 $\pm$   1.06 & &   2.66 $\pm$   0.12 &   3.80 $\pm$   0.24 & 4.00 \\
3798 H10                          &   3.11 $\pm$   0.05 &   5.56 $\pm$   0.13 & &   3.18 $\pm$   0.17 &   6.87 $\pm$   0.52 & &   3.92 $\pm$   0.18 &   5.29 $\pm$   0.29 & 5.34 \\
3820 He {\sc i}                   &   0.56 $\pm$   0.02 &   0.90 $\pm$   0.03 & &   0.11 $\pm$   0.08 &   0.12 $\pm$   0.10 & &   ...~~~~~~ &   ...~~~~~~ &   ...~~~~~~ \\
3835 H9                           &   4.66 $\pm$   0.07 &   7.91 $\pm$   0.15 & &   4.31 $\pm$   0.15 &   8.03 $\pm$   0.44 & &   5.84 $\pm$   0.13 &   7.57 $\pm$   0.23 & 7.37 \\
3869 [Ne {\sc iii}]               &  19.07 $\pm$   0.27 &  29.96 $\pm$   0.46 & &  27.21 $\pm$   0.45 &  31.38 $\pm$   0.56 & &  39.79 $\pm$   0.59 &  47.93 $\pm$   0.76 &   ...~~~~~~ \\
3889 He {\sc i} + H8              &  12.07 $\pm$   0.17 &  19.22 $\pm$   0.31 & &  11.65 $\pm$   0.24 &  16.54 $\pm$   0.48 & &  16.52 $\pm$   0.26 &  20.29 $\pm$   0.37 & 10.60 \\
3967 [Ne {\sc iii}]               &   6.08 $\pm$   0.17 &   9.08 $\pm$   0.26 & &   6.25 $\pm$   0.29 &   7.08 $\pm$   0.34 & &  11.05 $\pm$   0.17 &  13.04 $\pm$   0.21 &   ...~~~~~~ \\
3970 H7                           &  10.11 $\pm$   0.16 &  15.10 $\pm$   0.25 & &  11.20 $\pm$   0.34 &  12.69 $\pm$   0.40 & &  12.50 $\pm$   0.22 &  14.74 $\pm$   0.27 &16.00 \\
4026 He {\sc i}                   &   1.10 $\pm$   0.02 &   1.60 $\pm$   0.04 & &   0.74 $\pm$   0.17 &   0.82 $\pm$   0.20 & &   1.20 $\pm$   0.10 &   1.40 $\pm$   0.11 &   ...~~~~~~ \\
4068 [S {\sc ii}]                 &   1.52 $\pm$   0.03 &   2.15 $\pm$   0.05 & &   5.15 $\pm$   0.26 &   5.73 $\pm$   0.30 & &   0.84 $\pm$   0.06 &   0.98 $\pm$   0.07 &   ...~~~~~~ \\
4076 [S {\sc ii}]                 &   0.45 $\pm$   0.02 &   0.64 $\pm$   0.03 & &   ...~~~~~~ &   ...~~~~~~ & &   ...~~~~~~ &   ...~~~~~~ &   ...~~~~~~ \\
4101 H$\delta$                    &  19.53 $\pm$   0.28 &  27.69 $\pm$   0.42 & &  21.93 $\pm$   0.35 &  26.89 $\pm$   0.53 & &  22.61 $\pm$   0.35 &  26.36 $\pm$   0.44 & 26.10 \\
4177 [Fe {\sc ii}]                &   0.20 $\pm$   0.05 &   0.27 $\pm$   0.07 & &   ...~~~~~~ &   ...~~~~~~ & &   ...~~~~~~ &  ...~~~~~~ &   ...~~~~~~ \\
4233 Fe {\sc ii}                  &   0.18 $\pm$   0.04 &   0.23 $\pm$   0.05 & &   2.03 $\pm$   0.23 &   2.20 $\pm$   0.25 & &   ...~~~~~~ &   ...~~~~~~ &   ...~~~~~~ \\
4244 [Fe {\sc ii}]                &   0.29 $\pm$   0.03 &   0.38 $\pm$   0.04 & &   4.32 $\pm$   0.21 &   4.66 $\pm$   0.24 & &   ...~~~~~~ &   ...~~~~~~ &   ...~~~~~~ \\
4277 O {\sc ii}+[Fe {\sc ii}]     &   0.19 $\pm$   0.02 &   0.24 $\pm$   0.03 & &   3.44 $\pm$   0.21 &   3.69 $\pm$   0.23 & &   ...~~~~~~ &   ...~~~~~~ &   ...~~~~~~ \\
4287 [Fe {\sc ii}]                &   0.48 $\pm$   0.03 &   0.61 $\pm$   0.04 & &   4.80 $\pm$   0.20 &   5.14 $\pm$   0.22 & &   0.24 $\pm$   0.03 &   0.26 $\pm$   0.04 &   ...~~~~~~ \\
4340 H$\gamma$                    &  39.69 $\pm$   0.56 &  49.92 $\pm$   0.73 & &  40.96 $\pm$   0.62 &  46.08 $\pm$   0.77 & &  42.59 $\pm$   0.63 &  47.00 $\pm$   0.71 & 47.10 \\
4353 [Fe {\sc ii}]                &   0.05 $\pm$   0.02 &   0.06 $\pm$   0.03 & &   1.48 $\pm$   0.16 &   1.56 $\pm$   0.17 & &   ...~~~~~~ &   ...~~~~~~ &   ...~~~~~~ \\
4359 [Fe {\sc ii}]                &   0.53 $\pm$   0.05 &   0.66 $\pm$   0.06 & &   4.24 $\pm$   0.26 &   4.49 $\pm$   0.28 & &   ...~~~~~~ &   ...~~~~~~ &   ...~~~~~~ \\
4363 [O {\sc iii}]                &   1.98 $\pm$   0.04 &   2.45 $\pm$   0.05 & &   2.80 $\pm$   0.31 &   2.96 $\pm$   0.34 & &   8.80 $\pm$   0.15 &   9.59 $\pm$   0.17 &   ...~~~~~~\\
4387 He {\sc i}                   &   0.30 $\pm$   0.03 &   0.36 $\pm$   0.04 & &   ...~~~~~~ &   ...~~~~~~ & &   ...~~~~~~ &   ...~~~~~~ &   ...~~~~~~\\
4414 [Fe {\sc ii}]+O {\sc ii}     &   0.66 $\pm$   0.06 &   0.80 $\pm$   0.07 & &   4.87 $\pm$   0.52 &   5.11 $\pm$   0.56 & &   ...~~~~~~ &   ...~~~~~~ &   ...~~~~~~\\
4452 [Fe {\sc ii}]                &   0.28 $\pm$   0.05 &   0.34 $\pm$   0.06 & &   1.87 $\pm$   0.22 &   1.95 $\pm$   0.23 & &   ...~~~~~~ &   ...~~~~~~ &   ...~~~~~~\\
4458 [Fe {\sc ii}]                &   0.20 $\pm$   0.04 &   0.23 $\pm$   0.05 & &   ...~~~~~~ &   ...~~~~~~ & &   ...~~~~~~ &   ...~~~~~~ &   ...~~~~~~\\
4471 He {\sc i}                   &   3.40 $\pm$   0.06 &   4.00 $\pm$   0.07 & &   4.04 $\pm$   0.63 &   4.20 $\pm$   0.68 & &   3.22 $\pm$   0.09 &   3.44 $\pm$   0.10 &   ...~~~~~~\\
4475 [Fe {\sc ii}]                &   0.23 $\pm$   0.05 &   0.27 $\pm$   0.06 & &   3.24 $\pm$   0.30 &   3.37 $\pm$   0.32 & &   ...~~~~~~ &   ...~~~~~~ &   ...~~~~~~\\
4493 [Fe {\sc ii}]                &   0.20 $\pm$   0.03 &   0.23 $\pm$   0.04 & &   1.06 $\pm$   0.15 &   1.10 $\pm$   0.16 & &   ...~~~~~~ &   ...~~~~~~ &   ...~~~~~~\\
4515 [Fe {\sc ii}]                &   0.13 $\pm$   0.02 &   0.15 $\pm$   0.03 & &   ...~~~~~~ &   ...~~~~~~ & &   ...~~~~~~ &   ...~~~~~~ &   ...~~~~~~\\
4563 [Cr {\sc iii}]?                  &   0.09 $\pm$   0.01 &   0.10 $\pm$   0.02 & &   ...~~~~~~ &   ...~~~~~~ & &   ...~~~~~~ &   ...~~~~~~ &   ...~~~~~~\\
4571 Mg {\sc i}                     &   0.15 $\pm$   0.02 &   0.16 $\pm$   0.03 & &   ...~~~~~~ &   ...~~~~~~ & &   ...~~~~~~ &   ...~~~~~~ &   ...~~~~~~\\
4584 Fe {\sc ii}                  &   0.15 $\pm$   0.02 &   0.17 $\pm$   0.02 & &   1.84 $\pm$   0.15 &   1.88 $\pm$   0.16 & &   ...~~~~~~ &   ...~~~~~~ &   ...~~~~~~\\
4597 [Co {\sc iv}]?               &   0.09 $\pm$   0.02 &   0.10 $\pm$   0.02 & &   ...~~~~~~ &   ...~~~~~~ & &   ...~~~~~~ &   ...~~~~~~ &   ...~~~~~~\\
4607 N {\sc ii}+[Fe {\sc iii}]    &   0.14 $\pm$   0.02 &   0.15 $\pm$   0.03 & &   ...~~~~~~ &   ...~~~~~~ & &   ...~~~~~~ &   ...~~~~~~ &   ...~~~~~~\\
4630 N {\sc ii}                  &   0.30 $\pm$   0.03 &   0.33 $\pm$   0.04 & &   1.87 $\pm$   0.16 &   1.90 $\pm$   0.16 & &   ...~~~~~~ &   ...~~~~~~ &   ...~~~~~~\\
4640 N {\sc iii}                  &   0.19 $\pm$   0.04 &   0.21 $\pm$   0.04 & &   ...~~~~~~ &   ...~~~~~~ & &   ...~~~~~~ &   ...~~~~~~ &   ...~~~~~~\\
4658 [Fe {\sc iii}]               &   1.51 $\pm$   0.03 &   1.64 $\pm$   0.04 & &   3.36 $\pm$   0.21 &   3.39 $\pm$   0.21 & &   0.99 $\pm$   0.05 &   1.03 $\pm$   0.05 &   ...~~~~~~\\
4686 He {\sc ii}                  &   0.91 $\pm$   0.06 &   0.98 $\pm$   0.07 & &   ...~~~~~~ &   ...~~~~~~ & &   2.15 $\pm$   0.06 &   2.21 $\pm$   0.07 &   ...~~~~~~\\
4702 [Fe {\sc iii}]               &   0.34 $\pm$   0.03 &   0.36 $\pm$   0.04 & &   ...~~~~~~ &   ...~~~~~~ & &   0.26 $\pm$   0.04 &   0.26 $\pm$   0.04 &   ...~~~~~~\\
4712 [Ar {\sc iv}] + He {\sc i}   &   0.67 $\pm$   0.03 &   0.71 $\pm$   0.04 & &   ...~~~~~~ &   ...~~~~~~ & &   0.90 $\pm$   0.07 &   0.92 $\pm$   0.07 &   ...~~~~~~\\
4728 [Fe {\sc ii}]                &   0.16 $\pm$   0.03 &   0.17 $\pm$   0.03 & &   ...~~~~~~ &   ...~~~~~~ & &   ...~~~~~~ &   ...~~~~~~ &   ...~~~~~~\\
4740 [Ar {\sc iv}]                &   0.17 $\pm$   0.01 &   0.17 $\pm$   0.02 & &   ...~~~~~~ &   ...~~~~~~ & &   0.83 $\pm$   0.07 &   0.84 $\pm$   0.07 &   ...~~~~~~\\
4755 [Fe {\sc iii}]               &   0.36 $\pm$   0.03 &   0.37 $\pm$   0.03 & &   ...~~~~~~ &   ...~~~~~~ & &   ...~~~~~~ &   ...~~~~~~ &   ...~~~~~~\\
4770 [Fe {\sc iii}]               &   0.14 $\pm$   0.02 &   0.14 $\pm$   0.02 & &   ...~~~~~~ &   ...~~~~~~ & &   ...~~~~~~ &   ...~~~~~~ &   ...~~~~~~\\
4778 [Fe {\sc iii}]               &   0.15 $\pm$   0.03 &   0.15 $\pm$   0.03 & &   ...~~~~~~ &   ...~~~~~~ & &   ...~~~~~~ &   ...~~~~~~ &   ...~~~~~~\\
4814 [Fe {\sc ii}]                &   0.27 $\pm$   0.02 &   0.28 $\pm$   0.02 & &   2.06 $\pm$   0.16 &   2.03 $\pm$   0.16 & &   ...~~~~~~ &   ...~~~~~~ &   ...~~~~~~\\
4861 H$\beta$                     & 100.00 $\pm$   1.42 & 100.00 $\pm$   1.44 & & 100.00 $\pm$   1.45 & 100.00 $\pm$   1.50 & & 100.00 $\pm$   1.45 & 100.00 $\pm$   1.46 & 100.00 \\
4881 [Fe {\sc iii}]               &   0.47 $\pm$   0.02 &   0.47 $\pm$   0.02 & &   ...~~~~~~ &   ...~~~~~~ & &   0.53 $\pm$   0.04 &   0.53 $\pm$   0.04 &   ...~~~~~~\\
4890 [Fe {\sc ii}]                &   0.16 $\pm$   0.01 &   0.16 $\pm$   0.02 & &   1.14 $\pm$   0.10 &   1.11 $\pm$   0.10 & &   ...~~~~~~ &   ...~~~~~~ &   ...~~~~~~\\
4905 [Fe {\sc ii}]                &   0.13 $\pm$   0.01 &   0.13 $\pm$   0.01 & &   1.30 $\pm$   0.14 &   1.26 $\pm$   0.14 & &   0.25 $\pm$   0.02 &   0.25 $\pm$   0.02 &   ...~~~~~~\\
4922 He {\sc i}                   &   1.15 $\pm$   0.03 &   1.12 $\pm$   0.03 & &   ...~~~~~~ &   ...~~~~~~ & &   0.63 $\pm$   0.05 &   0.62 $\pm$   0.05 &   ...~~~~~~\\
4930 [Fe {\sc iii}]               &   0.20 $\pm$   0.03 &   0.19 $\pm$   0.03 & &   ...~~~~~~ &   ...~~~~~~ & &   ...~~~~~~ &   ...~~~~~~ &   ...~~~~~~\\
4943 O {\sc ii}                   &   0.09 $\pm$   0.02 &   0.09 $\pm$   0.02 & &   ...~~~~~~ &   ...~~~~~~ & &   ...~~~~~~ &   ...~~~~~~ &   ...~~~~~~\\
4959 [O {\sc iii}]                & 124.43 $\pm$   1.76 & 119.38 $\pm$   1.72 & & 102.66 $\pm$   1.49 &  99.02 $\pm$   1.48 & & 171.17 $\pm$   2.46 & 167.96 $\pm$   2.44 &   ...~~~~~~\\
4986 [Fe {\sc iii}]               &   0.61 $\pm$   0.02 &   0.58 $\pm$   0.02 & &   1.36 $\pm$   0.20 &   1.31 $\pm$   0.20 & &   1.25 $\pm$   0.05 &   1.22 $\pm$   0.05 &   ...~~~~~~\\
5007 [O {\sc iii}]                & 394.68 $\pm$   5.59 & 371.56 $\pm$   5.34 & & 331.11 $\pm$   4.76 & 317.19 $\pm$   4.68 & & 531.05 $\pm$   7.61 & 517.00 $\pm$   7.47 &   ...~~~~~~\\
5016 He {\sc i}                   &   1.38 $\pm$   0.03 &   1.29 $\pm$   0.03 & &   2.32 $\pm$   0.22 &   2.22 $\pm$   0.22 & &   1.78 $\pm$   0.06 &   1.73 $\pm$   0.06 &   ...~~~~~~\\
5041 Si {\sc ii}                  &   0.25 $\pm$   0.03 &   0.23 $\pm$   0.03 & &   ...~~~~~~ &   ...~~~~~~ & &   ...~~~~~~ &   ...~~~~~~ &   ...~~~~~~\\
5048 He {\sc i}                   &   0.13 $\pm$   0.02 &   0.12 $\pm$   0.02 & &   ...~~~~~~ &   ...~~~~~~ & &   ...~~~~~~ &   ...~~~~~~ &   ...~~~~~~\\
5056 Si {\sc ii}                  &   0.41 $\pm$   0.03 &   0.37 $\pm$   0.03 & &   ...~~~~~~ &   ...~~~~~~ & &   ...~~~~~~ &   ...~~~~~~ &   ...~~~~~~\\
5159 [Fe {\sc ii}]                &   0.65 $\pm$   0.02 &   0.58 $\pm$   0.02 & &   5.66 $\pm$   0.19 &   5.31 $\pm$   0.18 & &   ...~~~~~~ &   ...~~~~~~ &   ...~~~~~~\\
5169 Fe {\sc ii}                  &   0.14 $\pm$   0.02 &   0.12 $\pm$   0.01 & &   1.53 $\pm$   0.14 &   1.43 $\pm$   0.13 & &   ...~~~~~~ &   ...~~~~~~ &   ...~~~~~~\\
5199 [N {\sc i}]                  &   1.79 $\pm$   0.03 &   1.57 $\pm$   0.03 & &   4.83 $\pm$   0.17 &   4.50 $\pm$   0.17 & &   0.50 $\pm$   0.05 &   0.47 $\pm$   0.05 &   ...~~~~~~\\
5220 [Fe {\sc ii}]                &   0.14 $\pm$   0.04 &   0.12 $\pm$   0.03 & &   ...~~~~~~ &   ...~~~~~~ & &   ...~~~~~~ &   ...~~~~~~ &   ...~~~~~~\\
5235 Fe {\sc ii}?                 &   0.19 $\pm$   0.03 &   0.17 $\pm$   0.02 & &   1.28 $\pm$   0.13 &   1.19 $\pm$   0.13 & &   ...~~~~~~ &   ...~~~~~~ &   ...~~~~~~\\
5262 [Fe {\sc ii}]                &   0.29 $\pm$   0.02 &   0.25 $\pm$   0.01 & &   1.68 $\pm$   0.12 &   1.55 $\pm$   0.11 & &   ...~~~~~~ &   ...~~~~~~ &   ...~~~~~~\\
5270 [Fe {\sc iii}]               &   1.06 $\pm$   0.03 &   0.90 $\pm$   0.02 & &   2.82 $\pm$   0.12 &   2.60 $\pm$   0.12 & &   0.58 $\pm$   0.04 &   0.54 $\pm$   0.03 &   ...~~~~~~\\
5273 [Fe {\sc ii}]                &   0.22 $\pm$   0.01 &   0.18 $\pm$   0.01 & &   2.52 $\pm$   0.11 &   2.33 $\pm$   0.11 & &   ...~~~~~~ &   ...~~~~~~ &   ...~~~~~~\\
5317 Fe {\sc ii}?                 &   0.26 $\pm$   0.02 &   0.21 $\pm$   0.01 & &   3.26 $\pm$   0.18 &   2.99 $\pm$   0.17 & &   ...~~~~~~ &   ...~~~~~~ &   ...~~~~~~\\
5334 [Fe {\sc ii}]                &   0.21 $\pm$   0.03 &   0.17 $\pm$   0.03 & &   1.78 $\pm$   0.14 &   1.63 $\pm$   0.13 & &   ...~~~~~~ &   ...~~~~~~ &   ...~~~~~~\\
5363 [Ni {\sc iv}]                &   0.17 $\pm$   0.06 &   0.14 $\pm$   0.05 & &   1.23 $\pm$   0.26 &   1.12 $\pm$   0.24 & &   ...~~~~~~ &   ...~~~~~~ &   ...~~~~~~\\
5376 [Fe {\sc ii}]                &   0.20 $\pm$   0.03 &   0.17 $\pm$   0.02 & &   2.09 $\pm$   0.29 &   1.91 $\pm$   0.27 & &   ...~~~~~~ &   ...~~~~~~ &   ...~~~~~~\\
5518 [Cl {\sc iii}]               &   0.38 $\pm$   0.03 &   0.29 $\pm$   0.02 & &   0.86 $\pm$   0.13 &   0.77 $\pm$   0.12 & &   ...~~~~~~ &   ...~~~~~~ &   ...~~~~~~\\
5535 Fe {\sc ii}? N {\sc ii}? C {\sc ii}?      &   0.12 $\pm$   0.08 &   0.09 $\pm$   0.06 & &   ...~~~~~~ &   ...~~~~~~ & &   ...~~~~~~ &   ...~~~~~~ &   ...~~~~~~\\
5538 [Cl {\sc iii}]               &   0.26 $\pm$   0.02 &   0.20 $\pm$   0.01 & &   ...~~~~~~ &   ...~~~~~~ & &   ...~~~~~~ &   ...~~~~~~ &   ...~~~~~~\\
5755 [N {\sc ii}]                 &   1.62 $\pm$   0.03 &   1.16 $\pm$   0.02 & &   6.04 $\pm$   0.18 &   5.25 $\pm$   0.16 & &   ...~~~~~~ &   ...~~~~~~ &   ...~~~~~~\\
5876 He {\sc i}                   &  19.52 $\pm$   0.29 &  13.50 $\pm$   0.21 & &  16.12 $\pm$   0.47 &  13.81 $\pm$   0.42 & &  13.16 $\pm$   0.23 &  11.27 $\pm$   0.20 &   ...~~~~~~\\
5958 Si {\sc ii}                  &   0.16 $\pm$   0.02 &   0.08 $\pm$   0.01 & &   ...~~~~~~ &   ...~~~~~~ & &   ...~~~~~~ &   ...~~~~~~ &   ...~~~~~~\\
5979 Si {\sc ii}                  &   0.36 $\pm$   0.03 &   0.24 $\pm$   0.02 & &   ...~~~~~~ &   ...~~~~~~ & &   ...~~~~~~ &   ...~~~~~~ &   ...~~~~~~\\
6046 O {\sc i}                    &   0.13 $\pm$   0.04 &   0.09 $\pm$   0.02 & &   ...~~~~~~ &   ...~~~~~~ & &   ...~~~~~~ &   ...~~~~~~ &   ...~~~~~~\\
6300 [O {\sc i}]                  &   6.67 $\pm$   0.10 &   4.06 $\pm$   0.07 & &   9.55 $\pm$   0.27 &   7.81 $\pm$   0.23 & &   2.95 $\pm$   0.09 &   2.39 $\pm$   0.07 &   ...~~~~~~\\
6312 [S {\sc iii}]                &   1.82 $\pm$   0.05 &   1.10 $\pm$   0.03 & &   1.28 $\pm$   0.21 &   1.05 $\pm$   0.18 & &   1.62 $\pm$   0.08 &   1.32 $\pm$   0.06 &   ...~~~~~~\\
6347 Si {\sc ii}                  &   0.47 $\pm$   0.02 &   0.28 $\pm$   0.02 & &   ...~~~~~~ &   ...~~~~~~ & &   ...~~~~~~ &   ...~~~~~~ &   ...~~~~~~\\
6364 [O {\sc i}]                  &   2.36 $\pm$   0.04 &   1.41 $\pm$   0.02 & &   2.69 $\pm$   0.12 &   2.18 $\pm$   0.10 & &   0.66 $\pm$   0.10 &   0.53 $\pm$   0.08 &   ...~~~~~~\\
6365 [Ni {\sc ii}]                &   0.16 $\pm$   0.02 &   0.09 $\pm$   0.01 & &   ...~~~~~~ &   ...~~~~~~ & &   ...~~~~~~ &   ...~~~~~~ &   ...~~~~~~\\
6371 Si {\sc ii}                  &   0.30 $\pm$   0.02 &   0.18 $\pm$   0.01 & &   ...~~~~~~ &   ...~~~~~~ & &   ...~~~~~~ &   ...~~~~~~ &   ...~~~~~~\\
6384 Fe {\sc ii}                  &   0.19 $\pm$   0.04 &   0.11 $\pm$   0.02 & &   ...~~~~~~ &   ...~~~~~~ & &   ...~~~~~~ &   ...~~~~~~ &   ...~~~~~~\\
6456 Fe {\sc ii}                  &   0.25 $\pm$   0.04 &   0.15 $\pm$   0.02 & &   1.00 $\pm$   0.13 &   0.81 $\pm$   0.10 & &   ...~~~~~~ &   ...~~~~~~ &   ...~~~~~~\\
6516 Fe {\sc ii}                  &   0.12 $\pm$   0.02 &   0.07 $\pm$   0.01 & &   ...~~~~~~ &   ...~~~~~~ & &   ...~~~~~~ &   ...~~~~~~ &   ...~~~~~~\\
6548 [N {\sc ii}]                 &  28.91 $\pm$   0.41 &  16.38 $\pm$   0.26 & &  22.86 $\pm$   0.38 &  18.22 $\pm$   0.33 & &   2.91 $\pm$   0.07 &   2.29 $\pm$   0.06 &   ...~~~~~~\\
6563 H$\alpha$$^{\rm d}$                    & 514.82 $\pm$   7.29 & 290.71 $\pm$   4.53 & & 358.69$^{\rm d}$ $\pm$   5.17 & 286.36$^{\rm d}$ $\pm$   4.60 & & 374.18 $\pm$   5.37 & 294.56 $\pm$   4.62 &282.00 \\
6583 [N {\sc ii}]                 &  86.96 $\pm$   1.26 &  48.81 $\pm$   0.78 & &  72.57 $\pm$   1.06 &  57.64 $\pm$   0.94 & &   9.04 $\pm$   0.15 &   7.10 $\pm$   0.13 &   ...~~~~~~\\
6667 [Ni {\sc ii}]                &   0.18 $\pm$   0.03 &   0.10 $\pm$   0.02 & &   ...~~~~~~ &   ...~~~~~~ & &   ...~~~~~~ &   ...~~~~~~ &   ...~~~~~~\\
6678 He {\sc i}                   &   6.36 $\pm$   0.09 &   3.48 $\pm$   0.06 & &   3.50 $\pm$   0.10 &   2.75 $\pm$   0.08 & &   3.90 $\pm$   0.08 &   3.03 $\pm$   0.07 &   ...~~~~~~\\
6716 [S {\sc ii}]                 &  28.86 $\pm$   0.41 &  15.63 $\pm$   0.25 & &  37.16 $\pm$   0.55 &  29.14 $\pm$   0.48 & &  14.07 $\pm$   0.22 &  10.88 $\pm$   0.18 &   ...~~~~~~\\
6731 [S {\sc ii}]                 &  18.99 $\pm$   0.27 &  10.25 $\pm$   0.16 & &  17.08 $\pm$   0.27 &  13.37 $\pm$   0.24 & &  10.99 $\pm$   0.17 &   8.49 $\pm$   0.15 &   ...~~~~~~\\
7002 O {\sc i}                    &   0.09 $\pm$   0.01 &   0.04 $\pm$   0.00 & &   ...~~~~~~ &   ...~~~~~~ & &   ...~~~~~~ &   ...~~~~~~ &   ...~~~~~~\\
7065 He {\sc i}                   &   6.73 $\pm$   0.10 &   3.34 $\pm$   0.06 & &   3.98 $\pm$   0.09 &   3.02 $\pm$   0.07 & &   3.84 $\pm$   0.07 &   2.87 $\pm$   0.06 &   ...~~~~~~\\
7136 [Ar {\sc iii}]               &  14.37 $\pm$   0.21 &   7.01 $\pm$   0.12 & &  10.49 $\pm$   0.26 &   7.92 $\pm$   0.21 & &   7.56 $\pm$   0.13 &   5.60 $\pm$   0.11 &   ...~~~~~~\\
7155 [Fe {\sc ii}]                &   0.55 $\pm$   0.03 &   0.27 $\pm$   0.01 & &   1.76 $\pm$   0.22 &   1.32 $\pm$   0.17 & &   ...~~~~~~ &   ...~~~~~~ &   ...~~~~~~\\
7172 [Fe {\sc ii}]                &   0.14 $\pm$   0.02 &   0.07 $\pm$   0.01 & &   ...~~~~~~ &   ...~~~~~~ & &   ...~~~~~~ &   ...~~~~~~ &   ...~~~~~~\\
7236 C {\sc ii}                   &   0.19 $\pm$   0.01 &   0.09 $\pm$   0.01 & &   ...~~~~~~ &   ...~~~~~~ & &   ...~~~~~~ &   ...~~~~~~ &   ...~~~~~~\\
7254 O {\sc i}                    &   0.31 $\pm$   0.02 &   0.15 $\pm$   0.01 & &   ...~~~~~~ &   ...~~~~~~ & &   ...~~~~~~ &   ...~~~~~~ &   ...~~~~~~\\
7281 He {\sc i}                   &   0.94 $\pm$   0.02 &   0.44 $\pm$   0.01 & &   ...~~~~~~ &   ...~~~~~~ & &   0.68 $\pm$   0.04 &   0.50 $\pm$   0.03 &   ...~~~~~~\\
7291 [Ca {\sc ii}]                  &   0.19 $\pm$   0.02 &   0.09 $\pm$   0.01 & &   ...~~~~~~ &   ...~~~~~~ & &   ...~~~~~~ &   ...~~~~~~ &   ...~~~~~~\\
7320 [O {\sc ii}]                 &   5.32 $\pm$   0.08 &   2.49 $\pm$   0.04 & &   3.47 $\pm$   0.10 &   2.58 $\pm$   0.08 & &   2.42 $\pm$   0.06 &   1.76 $\pm$   0.04 &   ...~~~~~~\\
7330 [O {\sc ii}]                 &   4.38 $\pm$   0.06 &   2.04 $\pm$   0.03 & &   2.35 $\pm$   0.11 &   1.74 $\pm$   0.09 & &   2.08 $\pm$   0.05 &   1.51 $\pm$   0.04 &   ...~~~~~~\\
7378 [Ni {\sc ii}]                &   0.67 $\pm$   0.02 &   0.31 $\pm$   0.01 & &   1.66 $\pm$   0.12 &   1.23 $\pm$   0.09 & &   ...~~~~~~ &   ...~~~~~~ &   ...~~~~~~\\
7412 [Ni {\sc ii}]                &   0.15 $\pm$   0.02 &   0.07 $\pm$   0.01 & &   0.44 $\pm$   0.09 &   0.32 $\pm$   0.07 & &   ...~~~~~~ &   ...~~~~~~ &   ...~~~~~~\\
7711 Fe {\sc ii}                  &   0.08 $\pm$   0.01 &   0.03 $\pm$   0.00 & &   ...~~~~~~ &   ...~~~~~~ & &   ...~~~~~~ &   ...~~~~~~ &   ...~~~~~~\\
7751 [Ar {\sc iii}]               &   3.56 $\pm$   0.05 &   1.52 $\pm$   0.03 & &   1.30 $\pm$   0.10 &   0.94 $\pm$   0.07 & &   1.90 $\pm$   0.05 &   1.33 $\pm$   0.04 &   ...~~~~~~\\
7816 He {\sc i}                   &   0.10 $\pm$   0.02 &   0.04 $\pm$   0.01 & &   ...~~~~~~ &   ...~~~~~~ & &   ...~~~~~~ &   ...~~~~~~ &   ...~~~~~~\\
8125 [Cr {\sc ii}]                &   0.23 $\pm$   0.02 &   0.09 $\pm$   0.01 & &   ...~~~~~~ &   ...~~~~~~ & &   ...~~~~~~ &   ...~~~~~~ &   ...~~~~~~\\
8188 N {\sc i}                    &   0.37 $\pm$   0.03 &   0.14 $\pm$   0.01 & &   ...~~~~~~ &   ...~~~~~~ & &   ...~~~~~~ &   ...~~~~~~ &   ...~~~~~~\\
8216 N {\sc i}                    &   0.42 $\pm$   0.06 &   0.16 $\pm$   0.02 & &   ...~~~~~~ &   ...~~~~~~ & &   ...~~~~~~ &   ...~~~~~~ &   ...~~~~~~\\
8223 N {\sc i}?                   &   0.41 $\pm$   0.03 &   0.16 $\pm$   0.01 & &   ...~~~~~~ &   ...~~~~~~ & &   ...~~~~~~ &   ...~~~~~~ &   ...~~~~~~\\
8229 Fe {\sc ii}                  &   0.26 $\pm$   0.03 &   0.10 $\pm$   0.01 & &   ...~~~~~~ &   ...~~~~~~ & &   ...~~~~~~ &   ...~~~~~~ &   ...~~~~~~\\
8242 N {\sc i}                    &   0.27 $\pm$   0.02 &   0.10 $\pm$   0.01 & &   ...~~~~~~ &   ...~~~~~~ & &   ...~~~~~~ &   ...~~~~~~ &   ...~~~~~~\\
8299 P28                          &   0.09 $\pm$   0.01 &   0.03 $\pm$   0.00 & &   ...~~~~~~ &   ...~~~~~~ & &   ...~~~~~~ &   ...~~~~~~ &0.08 \\
8306 P27                          &   0.14 $\pm$   0.01 &   0.05 $\pm$   0.01 & &   ...~~~~~~ &   ...~~~~~~ & &   ...~~~~~~ &   ...~~~~~~ & 0.09 \\
8314 P26                          &   0.09 $\pm$   0.01 &   0.03 $\pm$   0.00 & &   ...~~~~~~ &   ...~~~~~~ & &   ...~~~~~~ &   ...~~~~~~ & 0.10 \\
8323 P25                          &   0.16 $\pm$   0.01 &   0.06 $\pm$   0.00 & &   ...~~~~~~ &   ...~~~~~~ & &   ...~~~~~~ &   ...~~~~~~ & 0.12 \\
8334 P24                          &   0.31 $\pm$   0.02 &   0.12 $\pm$   0.01 & &   ...~~~~~~ &   ...~~~~~~ & &   0.26 $\pm$   0.03 &   0.18 $\pm$   0.02 & 0.13 \\
8345 P23                          &   0.27 $\pm$   0.02 &   0.10 $\pm$   0.01 & &   ...~~~~~~ &   ...~~~~~~ & &   0.26 $\pm$   0.04 &   0.18 $\pm$   0.03 & 0.15 \\
8359 P22                          &   0.51 $\pm$   0.01 &   0.19 $\pm$   0.01 & &   ...~~~~~~ &   ...~~~~~~ & &   0.37 $\pm$   0.05 &   0.25 $\pm$   0.03 & 0.17 \\
8374 P21                          &   0.37 $\pm$   0.01 &   0.14 $\pm$   0.00 & &   ...~~~~~~ &   ...~~~~~~ & &   0.26 $\pm$   0.04 &   0.17 $\pm$   0.03 & 0.19 \\
8392 P20                          &   0.40 $\pm$   0.01 &   0.15 $\pm$   0.00 & &   ...~~~~~~ &   ...~~~~~~ & &   0.37 $\pm$   0.04 &   0.25 $\pm$   0.03 & 0.22 \\
8413 P19                          &   0.45 $\pm$   0.01 &   0.17 $\pm$   0.00 & &   ...~~~~~~ &   ...~~~~~~ & &   0.35 $\pm$   0.04 &   0.24 $\pm$   0.03 & 0.26 \\
8438 P18                          &   0.64 $\pm$   0.01 &   0.24 $\pm$   0.01 & &   ...~~~~~~ &   ...~~~~~~ & &   0.37 $\pm$   0.03 &   0.25 $\pm$   0.02 & 0.31 \\
8446 O {\sc i}                    &   3.93 $\pm$   0.06 &   1.47 $\pm$   0.03 & &   7.34 $\pm$   0.16 &   5.03 $\pm$   0.12 & &   0.74 $\pm$   0.04 &   0.49 $\pm$   0.03 &   ...~~~~~~\\
8470 P17                          &   0.62 $\pm$   0.01 &   0.23 $\pm$   0.01 & &   ...~~~~~~ &   ...~~~~~~ & &   0.36 $\pm$   0.04 &   0.24 $\pm$   0.03 & 0.36 \\
8490 Fe {\sc ii}                  &   0.19 $\pm$   0.01 &   0.07 $\pm$   0.01 & &   ...~~~~~~ &   ...~~~~~~ & &   ...~~~~~~ &   ...~~~~~~ &   ...~~~~~~\\
8505 P16                          &   0.76 $\pm$   0.02 &   0.28 $\pm$   0.01 & &   ...~~~~~~ &   ...~~~~~~ & &   0.41 $\pm$   0.03 &   0.27 $\pm$   0.02 & 0.44 \\
8548 P15                          &   0.90 $\pm$   0.02 &   0.33 $\pm$   0.01 & &   ...~~~~~~ &   ...~~~~~~ & &   0.37 $\pm$   0.04 &   0.24 $\pm$   0.02 & 0.53 \\
8579 ?                            &   0.31 $\pm$   0.02 &   0.11 $\pm$   0.01 & &   ...~~~~~~ &   ...~~~~~~ & &   ...~~~~~~ &   ...~~~~~~ &   ...~~~~~~\\
8601 P14                          &   1.25 $\pm$   0.02 &   0.46 $\pm$   0.01 & &   ...~~~~~~ &   ...~~~~~~ & &   0.70 $\pm$   0.04 &   0.46 $\pm$   0.03 & 0.65 \\
8617 [Fe {\sc ii}]                &   0.52 $\pm$   0.02 &   0.19 $\pm$   0.01 & &   ...~~~~~~ &   ...~~~~~~ & &   ...~~~~~~ &   ...~~~~~~ &   ...~~~~~~\\
8667 P13                          &   1.68 $\pm$   0.03 &   0.60 $\pm$   0.01 & &   ...~~~~~~ &   ...~~~~~~ & &   0.76 $\pm$   0.04 &   0.50 $\pm$   0.03 & 0.81 \\
8683 N {\sc i}                    &   0.48 $\pm$   0.02 &   0.17 $\pm$   0.01 & &   ...~~~~~~ &   ...~~~~~~ & &   ...~~~~~~ &   ...~~~~~~ &   ...~~~~~~\\
8753 P12                          &   2.19 $\pm$   0.04 &   0.78 $\pm$   0.02 & &   0.81 $\pm$   0.06 &   0.55 $\pm$   0.04 & &   1.23 $\pm$   0.08 &   0.80 $\pm$   0.05 & 1.04 \\
8865 P11                          &   2.81 $\pm$   0.04 &   0.98 $\pm$   0.02 & &   ...~~~~~~ &   ...~~~~~~ & &   1.62 $\pm$   0.05 &   1.04 $\pm$   0.03 & 1.35 \\
9017 P10                          &   3.66 $\pm$   0.06 &   1.24 $\pm$   0.02 & &   ...~~~~~~ &   ...~~~~~~ & &   2.13 $\pm$   0.05 &   1.35 $\pm$   0.04 & 1.80 \\
9069 [S {\sc iii}]                &  46.14 $\pm$   0.65 &  15.58 $\pm$   0.29 & &  21.14 $\pm$   0.33 &  14.43 $\pm$   0.29 & &  15.17 $\pm$   0.23 &   9.68 $\pm$   0.19 &   ...~~~~~~ \\
9095 Ca {\sc i}]                 &   0.19 $\pm$   0.02 &   0.06 $\pm$   0.01 & &   ...~~~~~~ &   ...~~~~~~ & &   ...~~~~~~ &   ...~~~~~~ &   ...~~~~~~\\
9231 P9                           &   6.11 $\pm$   0.10 &   2.01 $\pm$   0.04 & &   ...~~~~~~ &   ...~~~~~~ & &   1.42$^{\rm e}$ $\pm$   0.07 &   0.89$^{\rm e}$ $\pm$   0.04 & 2.49 \\
9464 He {\sc i}                   &   0.37 $\pm$   0.02 &   0.12 $\pm$   0.01 & &   ...~~~~~~ &   ...~~~~~~ & &   ...~~~~~~ &   ...~~~~~~ &   ...~~~~~~\\
9532 [S {\sc iii}]                & 136.72 $\pm$   1.94 &  43.16 $\pm$   0.82 & &  59.08 $\pm$   0.86 &  38.54 $\pm$   0.76 & &  41.96 $\pm$   0.62 &  25.96 $\pm$   0.50 &   ...~~~~~~\\
9548 Pe                           &  10.08 $\pm$   0.15 &   3.17 $\pm$   0.06 & &   3.26 $\pm$   0.12 &   2.10 $\pm$   0.08 & &   3.51 $\pm$   0.09 &   2.17 $\pm$   0.06 & 3.57 \\
9850 [C {\sc i}]                  &   0.57 $\pm$   0.06 &   0.17 $\pm$   0.02 & &   ...~~~~~~ &   ...~~~~~~ & &   ...~~~~~~ &   ...~~~~~~ &   ...~~~~~~\\ \\
  {}&\multicolumn{8}{c}{b) NIR range}\\ \\
10050 Pd                          &   ...~~~~~~ &   ...~~~~~~ & &   ...~~~~~~ &   ...~~~~~~ & &   9.56 $\pm$   0.32 &   5.75 $\pm$   0.21 & 5.40 \\
10432 [Fe {\sc ii}]?               &   0.18 $\pm$   0.06 &   0.05 $\pm$   0.02 & &   ...~~~~~~ &   ...~~~~~~ & &   ...~~~~~~ &   ...~~~~~~ &   ...~~~~~~\\
10502 Fe {\sc ii}                 &   0.40 $\pm$   0.04 &   0.11 $\pm$   0.01 & &   1.94 $\pm$   0.22 &   1.20 $\pm$   0.14 & &   ...~~~~~~ &   ...~~~~~~ &   ...~~~~~~\\
10551 ?                           &   0.27 $\pm$   0.03 &   0.07 $\pm$   0.01 & &   ...~~~~~~ &   ...~~~~~~ & &   ...~~~~~~ &   ...~~~~~~ &   ...~~~~~~\\
10715 [Ni {\sc ii}]               &   0.21 $\pm$   0.04 &   0.06 $\pm$   0.01 & &   ...~~~~~~ &   ...~~~~~~ & &   ...~~~~~~ &   ...~~~~~~ &   ...~~~~~~\\
10829 He {\sc i}                  & 120.40 $\pm$   1.71 &  32.64 $\pm$   0.66 & &  46.44 $\pm$   0.76 &  28.33 $\pm$   0.63 & &  47.85 $\pm$   2.23 &  27.75 $\pm$   1.36 &   ...~~~~~~\\
10910 He {\sc i}                  &   1.58 $\pm$   0.11 &   0.42 $\pm$   0.03 & &   ...~~~~~~ &   ...~~~~~~ & &   ...~~~~~~ &   ...~~~~~~ &   ...~~~~~~\\
10941 P$\gamma$                   &  20.45$^{\rm e}$ $\pm$   0.32 &   5.48$^{\rm e}$ $\pm$   0.12 & &   8.40$^{\rm e}$ $\pm$   0.38 &   5.10$^{\rm e}$ $\pm$   0.25 & &  13.90 $\pm$   0.25 &   8.03 $\pm$   0.18 & 8.77 \\
11287 O {\sc i}+Fe {\sc ii}       &   1.74 $\pm$   0.10 &   0.45 $\pm$   0.03 & &   ...~~~~~~ &   ...~~~~~~ & &   ...~~~~~~ &   ...~~~~~~ &   ...~~~~~~\\
11620 H$_2$              &0.18 $\pm$ 0.04 & 0.05 $\pm$ 0.01 &&   ...~~~~~~ &   ...~~~~~~ & &   ...~~~~~~ &   ...~~~~~~ &   ...~~~~~~\\
11850 H$_2$              &0.19 $\pm$ 0.04 & 0.05 $\pm$ 0.01 &&   ...~~~~~~ &   ...~~~~~~ & &   ...~~~~~~ &   ...~~~~~~ &   ...~~~~~~\\
11880 H$_2$+[Fe {\sc ii}]?        &   0.83 $\pm$   0.05 &   0.20 $\pm$   0.01 & &   ...~~~~~~ &   ...~~~~~~ & &   ...~~~~~~ &   ...~~~~~~ &   ...~~~~~~\\
11970 He {\sc i}                  &   0.51 $\pm$   0.05 &   0.13 $\pm$   0.01 & &   ...~~~~~~ &   ...~~~~~~ & &   ...~~~~~~ &   ...~~~~~~ &   ...~~~~~~\\
12070 H$_2$              &0.09 $\pm$ 0.03 & 0.02 $\pm$ 0.01 &&   ...~~~~~~ &   ...~~~~~~ & &   ...~~~~~~ &   ...~~~~~~ &   ...~~~~~~\\
12289 N {\sc i}                   &   0.15 $\pm$   0.03 &   0.04 $\pm$   0.01 & &   ...~~~~~~ &   ...~~~~~~ & &   ...~~~~~~ &   ...~~~~~~ &   ...~~~~~~\\
12330  H$_2$                   &   0.52 $\pm$   0.04 &   0.13 $\pm$   0.01 & &   ...~~~~~~ &   ...~~~~~~ & &   ...~~~~~~ &   ...~~~~~~ &   ...~~~~~~\\
12380 H$_2$+Fe {\sc ii}?           &   0.19 $\pm$   0.03 &   0.04 $\pm$   0.01 & &   ...~~~~~~ &   ...~~~~~~ & &   ...~~~~~~ &   ...~~~~~~ &   ...~~~~~~\\
12525 ?                           &   0.90 $\pm$   0.04 &   0.21 $\pm$   0.01 & &   ...~~~~~~ &   ...~~~~~~ & &   ...~~~~~~ &   ...~~~~~~ &   ...~~~~~~\\
12570 [Fe {\sc ii}]               &   6.14 $\pm$   0.11 &   1.44 $\pm$   0.03 & &   4.77 $\pm$   0.22 &   2.77 $\pm$   0.14 & &   ...~~~~~~ &   ...~~~~~~ &   ...~~~~~~\\
12790 He {\sc i}                  &   3.83 $\pm$   0.07 &   0.89 $\pm$   0.02 & &   2.77 $\pm$   0.19 &   1.59 $\pm$   0.12 & &   0.88 $\pm$   0.05 &   0.48 $\pm$   0.03 &   ...~~~~~~\\
12821 P$\beta$                    &  62.09 $\pm$   0.88 &  14.36 $\pm$   0.31 & &  30.54 $\pm$   0.53 &  17.59 $\pm$   0.42 & &  27.22 $\pm$   0.42 &  14.78 $\pm$   0.33 & 15.70 \\
13165 O {\sc i}                   &   0.92 $\pm$   0.10 &   0.21 $\pm$   0.02 & &   ...~~~~~~ &   ...~~~~~~ & &   ...~~~~~~ &   ...~~~~~~ &   ...~~~~~~\\
14967 Br25                        &   0.34 $\pm$   0.03 &   0.07 $\pm$   0.01 & &   ...~~~~~~ &   ...~~~~~~ & &   ...~~~~~~ &   ...~~~~~~ & 0.06 \\
15005 Br24                        &   0.28 $\pm$   0.03 &   0.06 $\pm$   0.01 & &   ...~~~~~~ &   ...~~~~~~ & &   ...~~~~~~ &   ...~~~~~~ & 0.06 \\
15043 Br23                        &   0.25 $\pm$   0.02 &   0.05 $\pm$   0.00 & &   ...~~~~~~ &   ...~~~~~~ & &   ...~~~~~~ &   ...~~~~~~ & 0.07 \\
15087 Br22                        &   0.34 $\pm$   0.03 &   0.07 $\pm$   0.01 & &   ...~~~~~~ &   ...~~~~~~ & &   ...~~~~~~ &   ...~~~~~~ & 0.08 \\
15137 Br21                        &   0.35 $\pm$   0.03 &   0.07 $\pm$   0.01 & &   ...~~~~~~ &   ...~~~~~~ & &   ...~~~~~~ &   ...~~~~~~ & 0.09 \\
15196 Br20                        &   0.50 $\pm$   0.03 &   0.10 $\pm$   0.01 & &   ...~~~~~~ &   ...~~~~~~ & &   ...~~~~~~ &   ...~~~~~~ & 0.11 \\
15265 Br19                        &   0.40 $\pm$   0.02 &   0.08 $\pm$   0.01 & &   ...~~~~~~ &   ...~~~~~~ & &   ...~~~~~~ &   ...~~~~~~ & 0.12 \\
15346 Br18                        &   1.19 $\pm$   0.04 &   0.24 $\pm$   0.01 & &   ...~~~~~~ &   ...~~~~~~ & &   ...~~~~~~ &   ...~~~~~~ & 0.15 \\
15443 Br17                        &   1.12 $\pm$   0.05 &   0.23 $\pm$   0.01 & &   ...~~~~~~ &   ...~~~~~~ & &   ...~~~~~~ &   ...~~~~~~ & 0.18 \\
15530 ?                           &   0.33 $\pm$   0.03 &   0.07 $\pm$   0.01 & &   ...~~~~~~ &   ...~~~~~~ & &   ...~~~~~~ &   ...~~~~~~ &   ...~~~~~~\\
15561 Br16                        &   1.05 $\pm$   0.07 &   0.21 $\pm$   0.01 & &   ...~~~~~~ &   ...~~~~~~ & &   ...~~~~~~ &   ...~~~~~~ & 0.21 \\
15705 Br15                        &   1.29 $\pm$   0.05 &   0.26 $\pm$   0.01 & &   ...~~~~~~ &   ...~~~~~~ & &   ...~~~~~~ &   ...~~~~~~ & 0.26 \\
15885 Br14                        &   1.49 $\pm$   0.04 &   0.30 $\pm$   0.01 & &   ...~~~~~~ &   ...~~~~~~ & &   0.44 $\pm$   0.05 &   0.22 $\pm$   0.03 & 0.31 \\
16010  H$_2$                   &   0.10 $\pm$   0.03 &   0.02 $\pm$   0.01 & &   ...~~~~~~ &   ...~~~~~~ & &   ...~~~~~~ &   ...~~~~~~ &   ...~~~~~~\\
16114 Br13                        &   2.08 $\pm$   0.04 &   0.42 $\pm$   0.01 & &   ...~~~~~~ &   ...~~~~~~ & &   0.71 $\pm$   0.04 &   0.36 $\pm$   0.02 & 0.39 \\
16410 Br12                        &   2.24 $\pm$   0.05 &   0.44 $\pm$   0.01 & &   ...~~~~~~ &   ...~~~~~~ & &   0.77 $\pm$   0.04 &   0.39 $\pm$   0.02 & 0.50 \\
16436 [Fe {\sc ii}]               &   6.28 $\pm$   0.10 &   1.25 $\pm$   0.03 & &   3.14 $\pm$   0.18 &   1.71 $\pm$   0.11 & &   ...~~~~~~ &   ...~~~~~~ &   ...~~~~~~\\
16640 [Fe {\sc ii}]               &   0.22 $\pm$   0.05 &   0.04 $\pm$   0.01 & &   ...~~~~~~ &   ...~~~~~~ & &   ...~~~~~~ &   ...~~~~~~ &   ...~~~~~~\\
16769 [Fe {\sc ii}]               &   0.36 $\pm$   0.03 &   0.07 $\pm$   0.01 & &   ...~~~~~~ &   ...~~~~~~ & &   ...~~~~~~ &   ...~~~~~~ &   ...~~~~~~\\
16811 Br11                        &   3.51 $\pm$   0.06 &   0.69 $\pm$   0.02 & &   ...~~~~~~ &   ...~~~~~~ & &   1.02 $\pm$   0.08 &   0.52 $\pm$   0.04 & 0.65 \\
16873 Fe {\sc ii}                 &   0.40 $\pm$   0.03 &   0.08 $\pm$   0.01 & &   ...~~~~~~ &   ...~~~~~~ & &   ...~~~~~~ &   ...~~~~~~ &   ...~~~~~~\\
17006 He {\sc i}                  &   1.16 $\pm$   0.03 &   0.23 $\pm$   0.01 & &   ...~~~~~~ &   ...~~~~~~ & &   0.55 $\pm$   0.04 &   0.28 $\pm$   0.02 &   ...~~~~~~\\
17367 Br10                        &   5.10 $\pm$   0.08 &   0.99 $\pm$   0.02 & &   2.60 $\pm$   0.37 &   1.41 $\pm$   0.20 & &   1.92 $\pm$   0.06 &   0.97 $\pm$   0.03 & 0.87 \\
18179 Br9                         &   7.44$^{\rm e}$ $\pm$   0.31 &   1.42$^{\rm e}$ $\pm$   0.06 & &   ...~~~~~~ &   ...~~~~~~ & &   ...~~~~~~ &   ...~~~~~~ & 1.21 \\
18693 ?                           &   9.90 $\pm$   0.34 &   1.88 $\pm$   0.07 & &   ...~~~~~~ &   ...~~~~~~ & &   ...~~~~~~ &   ...~~~~~~ &   ...~~~~~~\\
18756 P$\alpha$                   &  42.33$^{\rm e}$ $\pm$   0.77 &   8.02$^{\rm e}$ $\pm$   0.21 & &   ...~~~~~~ &   ...~~~~~~ & &  13.11$^{\rm e}$ $\pm$   0.23 &   6.55$^{\rm e}$ $\pm$   0.16 & 31.90 \\
19450 Br$\delta$                  &  10.06 $\pm$   0.17 &   1.89 $\pm$   0.05 & &   2.63 $\pm$   0.22 &   1.40 $\pm$   0.12 & &   1.18 $\pm$   0.09 &   0.59 $\pm$   0.05 & 1.73 \\
19540 He {\sc i}+[Fe {\sc ii}]    &   0.73 $\pm$   0.07 &   0.14 $\pm$   0.01 & &   ...~~~~~~ &   ...~~~~~~ & &   ...~~~~~~ &   ...~~~~~~ &   ...~~~~~~\\
19570 H$_2$                        &   0.61 $\pm$   0.03 &   0.11 $\pm$   0.01 & &   ...~~~~~~ &   ...~~~~~~ & &   0.58 $\pm$   0.07 &   0.29 $\pm$   0.03 &   ...~~~~~~\\
20340 H$_2$                        &   0.40 $\pm$   0.08 &   0.07 $\pm$   0.02 & &   ...~~~~~~ &   ...~~~~~~ & &   ...~~~~~~ &   ...~~~~~~ &   ...~~~~~~\\
20586 He {\sc i}                  &   8.22 $\pm$   0.13 &   1.52 $\pm$   0.04 & &   1.48 $\pm$   0.14 &   0.79 $\pm$   0.08 & &   2.02 $\pm$   0.09 &   1.00 $\pm$   0.05 &   ...~~~~~~\\
20730 H$_2$                       &   0.33 $\pm$   0.03 &   0.06 $\pm$   0.01 & &   ...~~~~~~ &   ...~~~~~~ & &   ...~~~~~~ &   ...~~~~~~ &   ...~~~~~~\\
20888 Fe {\sc ii}                 &   0.19 $\pm$   0.03 &   0.04 $\pm$   0.01 & &   ...~~~~~~ &   ...~~~~~~ & &   ...~~~~~~ &   ...~~~~~~ &   ...~~~~~~\\
21130 He {\sc i}                  &   0.57 $\pm$   0.04 &   0.11 $\pm$   0.01 & &   ...~~~~~~ &   ...~~~~~~ & &   ...~~~~~~ &   ...~~~~~~ &   ...~~~~~~\\
21220 H$_2$                       &   1.21 $\pm$   0.03 &   0.22 $\pm$   0.01 & &   ...~~~~~~ &   ...~~~~~~ & &  0.56 $\pm$0.06 & 0.28 $\pm$0.03&   ...~~~~~~\\
21620 He {\sc i}                  &   0.42 $\pm$   0.04 &   0.08 $\pm$   0.01 & &   ...~~~~~~ &   ...~~~~~~ & &   ...~~~~~~ &   ...~~~~~~ &   ...~~~~~~\\
21661 Br$\gamma$                  &  16.73 $\pm$   0.24 &   3.07 $\pm$   0.07 & &   5.38 $\pm$   0.19 &   2.85 $\pm$   0.12 & &   5.03 $\pm$   0.12 &   2.48 $\pm$   0.07 & 2.63 \\
22181 ?                           &   0.37 $\pm$   0.04 &   0.07 $\pm$   0.01 & &   ...~~~~~~ &   ...~~~~~~ & &   ...~~~~~~ &   ...~~~~~~ &   ...~~~~~~\\
22230 H$_2$                       &   0.26 $\pm$   0.05 &   0.05 $\pm$   0.01 & &   ...~~~~~~ &   ...~~~~~~ & &   ...~~~~~~ &   ...~~~~~~ &   ...~~~~~~\\ 
22480 H$_2$                       &   0.20 $\pm$   0.04 &   0.04 $\pm$   0.01 & &   ...~~~~~~ &   ...~~~~~~ & &   ...~~~~~~ &   ...~~~~~~ &   ...~~~~~~\\ \\
  {}&\multicolumn{8}{c}{c) MIR range$^{\rm f}$}\\ \\
10.51$\mu$m[S {\sc iv}]            &   12.76 $\pm$   0.10 &   ...~~~~~~ & &   ...~~~~~~ &   ...~~~~~~ & &   ...~~~~~~ &   ...~~~~~~ &   ...~~~~~~\\
12.81$\mu$m[Ne {\sc ii}]            &   9.07 $\pm$   0.14 &   ...~~~~~~  & &   ...~~~~~~ &   ...~~~~~~ & &   ...~~~~~~ &   ...~~~~~~ &   ...~~~~~~\\
15.55$\mu$m[Ne {\sc iii}]            &   28.51 $\pm$  0.40 &   ...~~~~~~  & &   ...~~~~~~ &   ...~~~~~~ & &   ...~~~~~~ &   ...~~~~~~ &   ...~~~~~~\\
18.71$\mu$m[S {\sc iii}]            &   13.14 $\pm$   0.12 &   ...~~~~~~  & &   ...~~~~~~ &   ...~~~~~~ & &   ...~~~~~~ &   ...~~~~~~ &   ...~~~~~~\\ \\
 $C$(H$\beta$) & \multicolumn{2}{c}{ 0.74 }& & \multicolumn{2}{c}{ 0.27 }& & \multicolumn{2}{c}{ 0.31 } & \\
 $F$(H$\beta$)$^{\rm g}$ & \multicolumn{2}{c}{344.20 }& & \multicolumn{2}{c}{ 63.59 }& & \multicolumn{2}{c}{121.40 } & \\
 EW(abs) \AA & \multicolumn{2}{c}{ 0.20 }& & \multicolumn{2}{c}{ 0.35 }& & \multicolumn{2}{c}{ 0.15 } &\\ \hline
  \end{longtable}
\noindent $^{\rm a}$ $F$($\lambda$) is observed flux, 
$F$($\lambda$)/$F$(H$\beta$) ratio designations with $\times$100. \\
$^{\rm b}$ $I$($\lambda$) is extinction-corrected flux,
$I$($\lambda$)/$I$(H$\beta$) ratio designations with $\times$100. \\
$^{\rm c}$ hydrogen recombination-line relative intensities by \citet{HS87}. \\
$^{\rm d}$ The observed flux of H$\alpha$ in Haro 11C is 
$F$(H$\alpha$)=3.350$\times$10$^{-14}$ erg s$^{-1}$ cm$^{-2}$. 
In this Table we show the $F$(H$\alpha$)/$F$(H$\beta$) ratio corresponding to 
$F$(H$\alpha$)=2.281$\times$10$^{-14}$ erg s$^{-1}$ cm$^{-2}$ 
obtained after subtraction of the P Cyg component. \\
$^{\rm e}$ affected by the telluric absorption. \\
$^{\rm f}$ $F$($\lambda$)/$F$(H$\beta$) ratio from \citet{Wu2008}.  \\
$^{\rm g}$ in units of 10$^{-16}$ erg s$^{-1}$ cm$^{-2}$.

\end{document}